\documentclass[journal]{IEEEtran}
\IEEEoverridecommandlockouts \makeatletter
\def\ps@headings{%
\def\@oddhead{\mbox{}\scriptsize\rightmark \hfil \thepage}%
\def\@evenhead{\scriptsize\thepage \hfil \leftmark\mbox{}}%
\def\@oddfoot{}%
\def\@evenfoot{}}
\makeatother
\usepackage[english]{babel}
\usepackage{amsmath,amsthm}
\usepackage{amsfonts}
\usepackage{amssymb}
\usepackage{enumerate}
\usepackage{enumitem}

\usepackage{graphicx}

\newtheorem*{construction}{VOA Construction}
\newtheorem*{remark}{Remark}
\newtheorem{question}{Question}

\usepackage{tikz}
\usepackage{multirow}
\usepackage{bm}
\usepackage{array}

\definecolor{orange-red}{rgb}{1.0, 0.27, 0.0}
\definecolor{darkred}{rgb}{0.55, 0.0, 0.0}

\usepackage{longtable}

\usepackage{soul}
\usepackage{color, xcolor}

\soulregister\cite7
\soulregister\ref7
\newcommand{\rv}[1]{{#1}}
\newcommand{\rvv}[1]{{#1}}

\newcommand{\mr}{\mathrm}
\newcommand{\mb}{\mathbf}
\newcommand{\mc}{\mathcal}

\newcommand{\h}{\mb{h}}

\newcommand{\br}{\mb{r}}

\newcommand{\ba}{\mb{a}}
\newcommand{\bb}{\mb{b}}
\newcommand{\bc}{\mb{c}}

\newcommand{\SetR}{\mathbb{R}}
\newcommand{\SetZ}{\mathbb{Z}}

\newcommand{\oa}{\mr{OA}}
\newcommand{\voa}{\mr{VOA}}
\newcommand{\mvoa}{\mr{MVOA}}
\newcommand{\ded}{\mr{ded}}

\newcommand{\Zv}{\mathbb{Z}_v}

\newtheorem{theorem}{Theorem}
\newtheorem{lemma}{Lemma}
\newtheorem{corollary}{Corollary}
\newtheorem{proposition}{Proposition}
\newtheorem{conjecture}{Conjecture}
\newtheorem{definition}{Definition}
\newtheorem{example}{Example}

\date{}

\begin{document}

\title{ {\huge Matroidal Entropy Functions: Constructions,
    Characterizations and Representations
  }
}

\author{Qi~Chen,
  Minquan~Cheng
and Baoming~Bai
\thanks{Qi Chen (qichen@xidian.edu.cn) is with School of
  Telecommunications, Xidian University, Xi’an 710071, China.

M. Cheng is with the Key Laboratory of Education Blockchain and Intelligent Technology, Ministry of Education,
and the Guangxi Key Laboratory of Multi-source Information Mining
\& Security, Guangxi Normal University, Guilin 541004, China (e-mail: chengqinshi@hotmail.com). 

    Baoming Bai (bmbai@mail.xidian.edu.cn) is with State Key Laboratory of Integrated Service
  Networks and School of Telecommunication Engineering, Xidian
  University, Xi’an 710071, China.

 This paper was in part presented at IEEE ISIT2022 \cite{CCB22}. }
}


\maketitle

\begin{abstract}
Matroidal entropy functions are entropy functions in the form $\h=\log v \cdot \br_M$, where $v\ge 2$
  is an integer and $\br_M$ is the rank function of a matroid
  $M$. They can be applied into capacity characterization and code
  construction of information theory problems such as network coding, secret
sharing, index coding and locally repairable code.
  In this paper, by
  constructing the variable strength orthogonal arrays of some matroid
  operations, we characterize matroidal entropy functions induced by
  regular matroids and some matroids with the same p-characteristic
  set as uniform matroid $U_{2,4}$. 

  Index terms-entropy function, matroid, matroidal entropy function,
  information inequalities, orthogonal array, 
  polymatroid, variable strength orthogonal array

\end{abstract}

\section{Introduction}
\label{sec:1}
For a discrete random vector $X_N\triangleq (X_i:i\in N)$, where $N$
is a finite set with cardinality $n$
, we define its
entropy function $\h_{X_N}: 2^N\to \SetR$ by
\begin{equation*}
  \h_{X_N}(A)=H(X_A) \quad \forall A\subseteq N,
\end{equation*}
where $X_A=(X_i:i\in A)$ and $H(X_A)$ is its joint entropy. We consider it as a vector in the entropy space
$\mc{H}_N\triangleq \SetR^{2^N}$ and denote the set of all entropy
functions in $\mc{H}_N$ by $\Gamma^*_N$. For each $\h\in\mc{H}_N$, we
call it \emph{entropic} if $\h\in \Gamma^*_n$.
The characterization of
entropy functions, i.e., determining whether a vector in the entropy
space
is entropic or not, is of fundamental importance in information
theory. 

In 1978, Fujishige proved in \cite{F78} that \rv{every} entropy function is (the rank function of) a polymatroid,
that is, a set function $\h : 2^N\to \SetR$ which satisfies
\begin{align*}
  \text{ (nonnegativity) } &\h(A)\ge 0, \quad \text{ for all } A\subseteq N,\\
  \text{ (monotonicity) } &\h(A)\le \h(B), \quad\text{ for all } A\subseteq
B\subseteq N,\\
  \text{(submodularity) } &\h(A\cap B)+\h(A\cup B)\le
\h(A)+\h(B), \\ &\quad\quad\quad \quad\quad\quad\text{ for all } A,B\subseteq N.
\end{align*}
We denote the set of all polymatroids in $\mc{H}_N$ by
$\Gamma_N$ and call the inequalities implied by those bounding
$\Gamma_N$ listed above \emph{Shannon-type} information inequalities,
as they correspond to the nonnegativity of the Shannon information
measures. \rv{So a vector $\h\in\mc{H}_N$ is a polymatroid if and only
  if it satisfies all Shannon-type inequalities.} In 1998, Zhang and Yeung discovered the first non-Shannon-type
information inequalities \cite{ZY98}, which implies that $\overline{\Gamma^*_N}$,
the closure of $\Gamma^*_N$, is a proper subset of $\Gamma_N$ when
$|N|\ge 4$. From then on, a series of non-Shannon-type information
inequalities were discovered, e.g., \cite{YYZ01,MMRV02,Z03,DFZ06}. In 2007, Mat\'u\v{s} proved in \cite{M07a} that for fixed
$N$ with $|N|\ge 4$, there exist infinitely many independent information
inequalities. Thus, to characterize entropy functions, we need to find ways other than information
inequalities.


\rv{Note that each non-redundant information inequality rules those
vectors in $\mc{H}_N$ not satisfying it out of the candidates
of entropy functions, and thus 
determines a tighter outer bound on $\Gamma^*_N$. In \cite{CCB21} and this paper, on
the contrary, we ensure previously unknown vectors in the
entropy space to be
entropic, and thus obtain a tighter inner bound on $\Gamma^*_N$. We do it from the 
boundary of $\Gamma_N$. }
We characterized the matroidal entropy functions \rv{which are on the
extreme rays, or $1$-dimensional faces of $\Gamma_n$ if it is induced
by a connected matroid (see Subsection \ref{extray} for more details).} For a
random vector $X_N$, if its entropy function is in the form
\begin{equation*}
  \h_{X_N}=\log v\cdot \br_M,
\end{equation*}
where $v\ge2$ is an integer and $\br_M$ is the rank function of a
matroid $M$ \rv{defined on the ground set $N$}(cf. Subsection
\ref{matroids}), we call  $\h_{X_N}$
a \emph{matroidal entropy function}. For a matroid $M$ with rank
function $\br_M$, the set
\begin{equation*}
  \chi_M\triangleq \{v\in \SetZ: v\ge 2,\ \log v\cdot \br_M \in\Gamma^*_N\}
\end{equation*}
is called the \emph{probabilistically characteristic
  set} \rv{(written as p-characteristic set for short)} of $M$.

In \cite[Theorem 1]{CCB21}, we proved that for a connected
matroid $M$ with rank exceeding 1 and integer $v\ge 2$, $\h=\log
v\cdot \br_M$ is in $\Gamma^*_N$ if and only if a variable strength orthogonal
arrays(VOA) $\voa(M,v)$ is constructible (cf. Subsection
\ref{voa} ). Thus, to characterize $\h=\log v\cdot \br_M$, or to
determine $\chi_M$ is equivalent to determine the set of all $v\ge 2$
such that a $\voa(M,v)$ exists.


For a $\voa(M,v)$ $\mb{T}$, let $\Omega$ be the set of all rows of
$\mb{T}$. It can be verified that for each $A\subseteq N$, if we put
all rows with the same sub-row indexed by $A$ in a block, we obtain a
partition $\xi_A$ of $\Omega$ with $v^{\br(A)}$ blocks all of the same
cardinality. The system of the partitions $\{\xi_A:A\subseteq N\}$ is
called a partition representation of $M$ with degree $v$
\cite{M99}. The partition representation problem of a matroid, which
is equivalent to the characterization of matroidal entropy function problem, is
an extension of the matroid representation problem over a field $\mathbb{F}$
\cite[Chapter 6]{O11}. The set $\Omega$ of all rows of
$\mb{T}$ also forms the code book of an $(M,v)$ almost affine
code \cite{SA98}. Thus, it can be applied into some coding problems such as secret
sharing\cite{BD91}, network coding\rv{\cite{DFZ04,DFZ07,SYLL15,LS01,TS19}}, index coding
\rv{\cite{RSG10,MS16}} and locally repairable code \cite{WFEH16}.

\rvv{Matroid theory not only provides an abundance of tools for characterizing entropy functions, but it also aids in understanding the relationships among matroidal entropy functions.}
In this paper, we utilize the correspondences among matroidal entropy
functions, matroids and VOAs developed in \cite{CCB21}  to
further characterize matroidal entropy functions. We construct
the VOAs of some matroid operations, including unitary operations
such as deletion, contraction and minor, and binary operations such as
parallel connection,
series connection and 2-sum, and determine the p-characteristic sets
of them. \rv{The minor structures among matroids imply the inclusion
  relations between their p-characteristic set. }
Armed with the tools \rv{we built up and using the minor structures of
matroids}
, we characterize matroidal entropy
functions induced by two families of matroids, i.e, regular
matroids\rv{, the family $\mc{M}_{\mr{reg}}$ of all matroids with
  p-characteristic set $\{v\in \SetZ: v\ge 2\}$, }
and matroids with the same p-characteristic set as $U_{2,4}$, \rv{the
  family $\mc{M}_{U_{2,4}}$ of all matroids with p-characteristic set $\{v\in \SetZ: v\ge
  3, v\neq 6\}$}. \rv{It is natural to define an equivalence relation on
  the set $\mc{M}$ of all matroids according to their p-characteristic
set. We say two matroids are p-characteristically equivalent if they
have the same  p-characteristic set. As we discuss in Subsection \ref{extrac},
  families $\mc{M}_{\mr{reg}}$ and $\mc{M}_{U_{2,4}}$ are the lowest
  two equivalence classes of
  matroids in the partially ordered set $(\mc{M}, \preceq)$ ordered by minor.}

{
\scriptsize 
\begin{table*}[htbp]
  \caption{Notation List}
\scalebox{0.85}{
  \begin{tabular}{lllc}
    \hline
\hline
$N$&$\{1,\cdots,n\}$& index set & Section \ref{sec:1}\\

$\mc{H}_N$&$\SetR^{2^N}$& entropy space \rvv{indexed by $N$}& \\

\rvv{$\Gamma^*_N$}& &entropy region \rvv{indexed by $N$}&\\

\rvv{$\Gamma_N$}& &polymatroidal region \rvv{indexed by $N$}&\\
$\chi_M$& &p-characteristic set of a matroid $M$& \\
 \hline
$M$ & $(N,\br)$ & matroid with ground set $N$ and rank function $\br$ &Subsection \ref{matroids}\\
& $(N,\br)$ & matroid with ground set $N$ and rank function $\br$ &\\
 & $(N,\mc{I})$ & matroid with ground set $N$ and family $\mc{I}$ of
                  independent sets& \\
 & $(N,\mc{B})$ & matroid with ground set $N$ and family $\mc{B}$ of
                  basis & \\
 & $(N,\mc{C})$ & matroid with ground set $N$ and family $\mc{C}$ of
                  circuits &\\

  $U_{t,n}$&  & uniform matroid with ground set $N$ and rank $t$ &\\
  \hline
$\Zv$ & $\{0,1,\ldots, v-1\}$ & set of symbols of an OA or a VOA & Subsection  \ref{voa}\\
& & a ring over $\Zv$ with usual addition and multiplication mod $v$  & \\
$\oa(\lambda\times v^t;t,n,v)$& & orthogonal array with index
                                  $\lambda$, strength $t$, factor $n$
                                  and level $v$ &\\
  $\oa(t,n,v)$& & orthogonal array with index
                                  unit, strength $t$, factor $n$
                  and level $v$ &\\
    $\voa(M,v)$& &variable strength orthogonal array induced by
                   matroid $M$ and with
                   level $v$ &\\
  \hline
  $\ded(\mb{T})$ & &deduplication,   the array
whose rows are those of $\mb{T}'$ with each occurring exactly
once& Subsection\ref{dcm} \\ 
 $M\setminus S$ , \rv{$M\setminus p$} \footnotemark{}  & & deletion of
                                                           $S$ from
                                                           $M$ &\\
    & & restriction of M on $N\setminus S$ &\\
  $\mb{T}\setminus S$ \rv{, $\mb{T}\setminus p$}& & deletion of $S$ from $\mb{T}$ & \\
 $M/ S$\rv{, $M/p$} & & a matroid of $M$ contracted by $S$ & \\
  $\mb{T}|_{S:\mb{a}}$\rv{, $\mb{T}|_{p:\mb{a}}$}& & contraction of $S$ from $\mb{T}$ to $\mb{a}$ & \\
  \hline

      $S(M_1,M_2)$  & $S((M_1;p_1),(M_2;p_2))$& series connection of
                                 matroids $M_1$ and $M_2$ with respect
                                              to base points $p_1$ and
                                 $p_2$ & Subsection\ref{bop}\\
      $P(M_1,M_2)$  & $P((M_1;p_1),(M_2;p_2))$& parallel connection of
                                 matroids $M_1$ and $M_2$ with respect
                                              to base points $p_1$ and
                                                $p_2$ & \\
    $S(\mb{T}_1,\mb{T}_2)$  & $S((\mb{T}_1;p_1),(\mb{T}_2;p_2))$& series connection of
                                VOA $\mb{T}_1$ and $\mb{T}_2$ with respect
                                              to base points $p_1$ and
                                 $p_2$ & \\
  
      $P(\mb{T}_1,\mb{T}_2)$  & $P((\mb{T}_1;p_1),(\mb{T}_2;p_2))$& parallel connection of
                                 VOA $\mb{T}_1$ and $\mb{T}_2$ with respect
                                              to base points $p_1$ and
                                                                    $p_2$
                                 & \\
  \hline
  
    $M_1\oplus M_2$  & & direct sum ($1$-sum ) of matroids $M_1$ and $M_2$ &  Subsection\ref{12sum}\\
  $\mb{T}_1\oplus \mb{T}_2$  & &  direct sum  ($1$-sum ) of VOAs $\mb{T}_1$ and
                                  $\mb{T}_2$ & \\
  $M_1\oplus_2 M_2$  & &  $2$-sum of matroids $M_1$ and $M_2$ & \\
  $\mb{T}_1\oplus_2\mb{T}_2$  & &  $2$-sum of matroids $\mb{T}_1$ and
                                  $\mb{T}_2$ & \\
  \hline
  $F_7$  & & Fano matroid & Subsection \ref{srm}\\
  $F^*_7$& & dual of Fano matroid &  \\
  \hline 
  $\mc{W}_r$  & & wheel graph with $r$ spokes & Subsection
                                                \ref{chu24}\\
    $M(\mc{W}_r)$  & & wheel matroid with rank $r$ & \\
  $\mc{W}^r$  & & whirl matroid with rank $r$ &\\
  \hline
  $P$ & $(N,\mb{r})$ & integer polymatroid with ground set $N$ and
                       rank function $\br$ &  Subsection \ref{extrac} \\
  $\mvoa(P, v)$ & & multi-level variable strength orthogonal array induced by
                   matroid $M$ and with base
                    level $v$ &  \\

\hline
\hline

  \end{tabular}
}

  \centering
  \label{tab:1}

\end{table*}

\footnotetext{\rv{When $S=\{p\}$, singleton, we write $M\setminus S$,
    $\mb{T}\setminus $, $M/ S$ and $\mb{T}|_{S:\mb{a}}$ as $M\setminus
    p$, $\mb{T}\setminus p$, $M/p$ and $\mb{T}_{p:\mb{a}}$,
    respectively for simplicity.}}
}

The rest of this paper is organized as follows. 
Section \ref{pre} gives the preliminaries on matroids and VOAs. In Section \ref{con}, we introduce how the p-characteristic set
$\chi_M$ of a matroid $M$ can be determined by the constructions of
matroids. Using the tools built up in  Section \ref{con}, we characterize matroidal entropy
functions induced by two families of matroids in Section \ref{pst}.
In
Section \ref{disc}, we discuss five topics about further research of
matroidal entropy functions and ask five questions to be answered, 
togheter with the applications of matroidal entropy functions to
network coding.
We close this section with the list of
notations in this paper in Table \ref{tab:1}.

\section{Preliminaries}
\label{pre}

\subsection{Matroids}
\label{matroids}
There exist a number of cryptomorphic definitions of a matroid. In
view of its relations to entropy functions, in this paper,
we define it by its rank function. 

\begin{definition}[\rv{\cite{O11}}]
  A \emph{matroid} $M$ is an ordered pair $(N, \mb{r})$, where the \emph{ground set}
        $N$ is a finite set and the \emph{rank function} $\mb{r}$ is a
        set function on $2^N$, and they satisfy the conditions
        that, for any
        $A,B\subseteq N$,
     \begin{itemize}
        \item $0\le\mb{r}(A)\le|A|$ \text{and}
        $\mb{r}(A)\in\SetZ$,
        \item $\mb{r}(A)\leq \mb{r}(B),\ \text{if }A\subseteq B$,
        \item $\mb{r}(A)+\mb{r}(B) \geq \mb{r}(A\cup B)+\mb{r}(A\cap B)$.
        \end{itemize}
\end{definition}
The value $\mb{r}(N)$ is called the \emph{rank}
of $M$. 

For the matroid $M=(N, \br)$, a subset $I\subseteq N$ is \rv{said to be} 
\emph{independent} if $|I|=\br(I)$. The family $\mc{I}$ of all
independent \rv{sets} is monotone, i.e., if $I_1\in \mc{I}$ and
$I_2\subseteq I_1$, then $I_2\in \mc{I}$. A subset of $N$ which is not 
independent is \rv{said to be} \emph{dependent}.
A maximal independent set
$B$ is called a \emph{basis} of $M$. Let $\mc{B}$ be the family of all
bases of $M$.
According to \cite[Lemma
1.2.1]{O11}, all members of $\mc{B}$ have the same cardinality as
$\mb{r}(N)$.
A subset $C\subseteq N$
is called a \emph{circuit} of $M$ if $C$ is dependent and for any
$e\in C$, $C-e$ is independent, that is, $C$ is a minimal dependent
set of $M$. Let $\mc{C}$ be the family of all
circuits of $M$. For $e\in N$, if $\{e\}\in \mc{C}$, it is called a
\emph{loop} and if for all $C\in \mc{C}$, $e\not\in C$, $e$ is called
a \emph{coloop}.
According to \cite[Theorem 1.3.2, 1.2.3, 1.1.4]{O11}, each of
$\mc{I}$, $\mc{B}$ and $\mc{C}$ uniquely specifies $M$. Thus, in the
following, we also say $M$ is a matroid $(N,\mc{I})$, $(N,\mc{B})$ or
$(N,\mc{C})$. 

For a matrix $\mb{A}$ over a field $\mathbb{F}$, let $N$ be the index set of
its columns, and $\mc{I}$ be the family of those sets of columns that
are linearly independent. By \cite[Proposition 1.1.1]{O11},
$M[\mb{A}]\triangleq (N,\mc{I})$
is the matroid with the family of independent
set $\mc{I}$. The matroid $M[\mb{A}]$ is called the \emph{vector matroid} of $A$. For a matroid $M$, if there exists a
matrix $\mb{A}$ over $\mathbb{F}$, \rv{and} $M$ is the vector matroid of $\mb{A}$, we say
that $M$ is $\mathbb{F}$-representable.

For a graph $G$ with the edge set $N$ and the family $\mc{C}$ of
cycles, by \cite[Proposition 1.1.7]{O11}, $M(G)\triangleq (N, \mc{C})$
is a matroid with ground set $N$ and family of circuits $\mc{C}$. The
matroid $M(G)$ is called the \emph{cycle matroid} of $G$.

For a matroid $M=(N,\mc{C})$, if for each two elements $e_1,e_2\in N$,
there exists a circuit $C\in \mc{C}$ such that $e_1,e_2\in C$, then
$M$ is called \emph{connected}. Note that for a graphical matroid $M$, if it
is connected, it is a cycle matroid of a $2$-connected graph without
loops. Thus, when we say a connected matroid in this paper, it is a $2$-connected
matroid \cite[Chapter 4]{O11}. In the following sections of this paper, we will also discuss
$3$-connected matroid and its relation to the matroid operation
$2$-sum. For the definition of hyper-connectivity of a matroid and its
properties, readers are
referred to \cite[Chapter 8]{O11}. 

\begin{example}
\label{asdf}
  Let $M=(N,\mb{r})$ be a matroid with $N=\{1,2,3,4,5,6\}$ and rank function
  \begin{equation*}
    \mb{r}(A)=
    \begin{cases}
     \  |A| \quad & |A|\leq 2,\\
     \  2\quad & A=\{4,5,6\},\\
     \  3\quad & |A|=3, A\neq \{4,5,6\} \\ & \text{ or }
     A=\{4,5,6\}\cup\{i\}, i=1,2,3 \\ &\text{ or } A=\{1,2,3,4\}, \\
     \  4\quad & \text{o.w.}
    \end{cases}
  \end{equation*}
 It can be checked that for this \rv{connected} matroid,
 $\mc{C}=\{\{1,2,3,4\},\{4,5,6\},\{1,2,3,5,6\}\}$, $\mc{B}=\{A\subseteq
 N: |A|=4, A\neq \{1,2,3,4\}, \{4,5,6\}\nsubseteq A\}$ and
 $\mc{I}=\{A\subseteq N: |A|\le 2\}\cup \{A\subseteq N: |A|=3, A\neq
 \{4,5,6\} \}\cup\mc{B}$. It is the vector matroid of the following
 matrix
 \begin{equation*}
   \begin{bmatrix}
     1 & 0 & 0 & 1 & 0 & 1 \\
     0 & 1 & 0 & 1 & 0 & 1 \\
     0 & 0 & 1 & 1 & 0 & 1 \\
     0 & 0 & 0 & 0 & 1 & 1   
   \end{bmatrix}
 \end{equation*}
 over $\mr{GF}(2)$ and cycle matroid of the graph in Fig \ref{m1}.

 \begin{figure}[h]
       \centering
       \includegraphics[width=6cm]{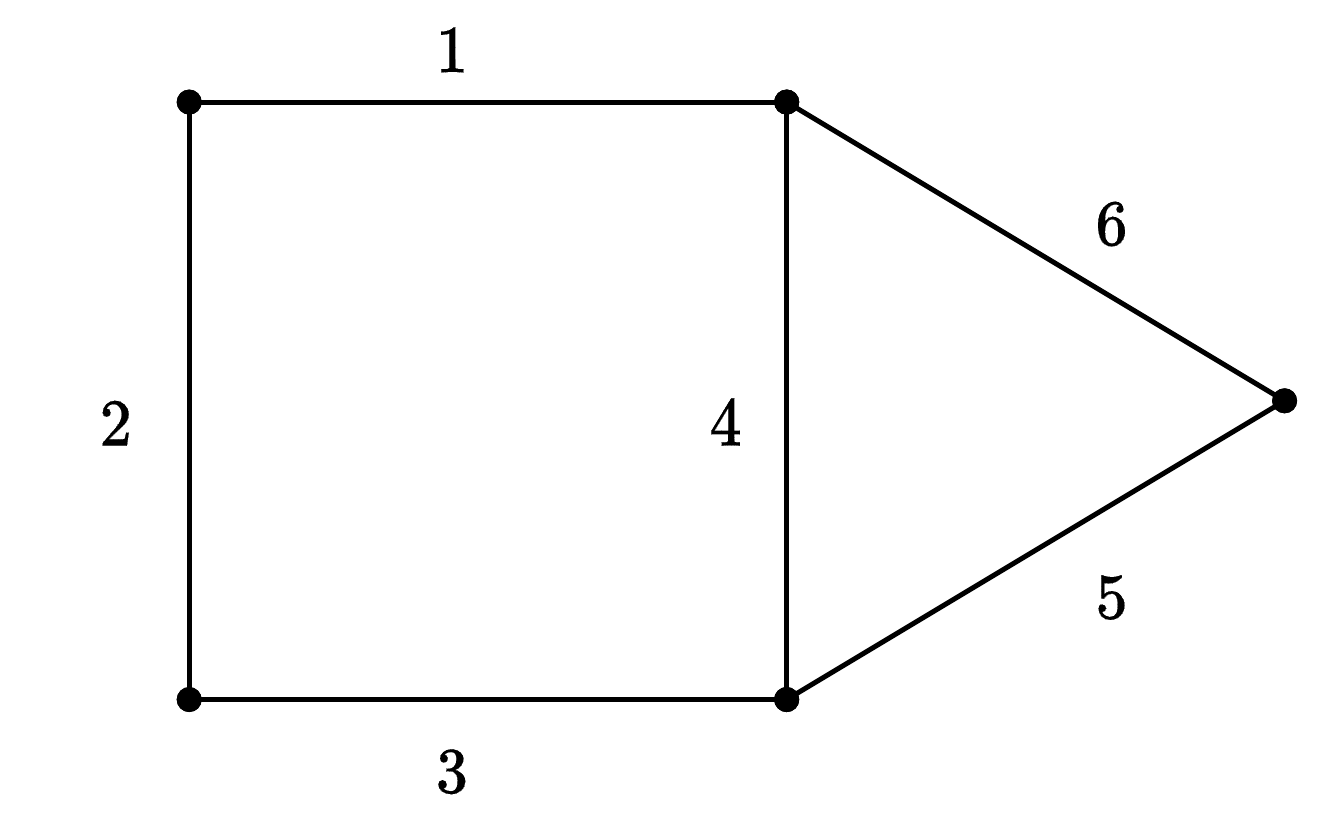}
       \caption{The graph whose cycle matroid is $M$ \rv{in Exmaple \ref{asdf}}.}
       \label{m1}
\end{figure}

\end{example}

\begin{example}

A matroid $U_{t,n}=(N,\br)$ with $n=|N|$, $0\le t\le n$ and
a rank function
\begin{equation*}
  \br(A)=\min\{t,|A|\}\quad A\subseteq N
\end{equation*}
is called uniform. For a uniform matroid, $\mc{I}=\{A\subseteq N:
|A|\le t\}$, $\mc{C}=\{A\subseteq N:
|A|=t+1\}$ and $\mc{B}=\{A\subseteq N:
|A|=t\}$. 
Uniform matroids play an important role in the
characterizations of matroidal entropy functions.  
\end{example}

\subsection{Variable strength orthogonal array}
\label{voa}
In \cite{CCB21}, we extend a well-studied structure in combinatorial
design theory orthogonal array(OA) to variable strength orthogonal array(VOA). 
Let  $\Zv\triangleq \{0,1,\ldots, v-1\}$ be the set of symbols of an OA or a VOA.

\begin{definition}[\rv{\cite{HSS99}}]
  A $\lambda v^t\times n$ array $\mb{T}$ with entries from
  $\Zv$ is called an \emph{orthogonal
  array} of \emph{strength} $t$, \emph{factor} $n$, \emph{level} $v$
and \emph{index} $\lambda$ if for each
  $\lambda v^t\times t$ subarray $\mb{T}'$ of
  $\mb{T}$, each $t$-tuple in $\Zv^t$ occurs in the rows of $\mb{T}'$ exactly $\lambda$ times. We
  call $\mb{T}$ an $\oa(\lambda\times v^t;t,n,v)$. When $\lambda=1$,
  we say such orthogonal array has \emph{index unity} and call it an
  $\oa(t,n,v)$ for short.
\end{definition}

\begin{definition}[\rv{\cite{CCB21}}]
  \label{dvoa}
  Given a loopless matroid $M=(N,\mb{r})$ with $\br(N)\ge 2$, a $v^{\br(N)}\times n$
  array $\mb{T}$ with columns indexed by $N$, entries from $\Zv$, is called a variable strength
  orthogonal array(VOA) induced by $M$ with level $v$ if for each
 $A\subseteq N$, \rv{the} $v^{\mb{r}(N)}\times |A|$ subarray $\mb{T}(A)$ of $\mb{T}$ consisting of
 columns indexed by $A$ satisfies the following condition:
each row of  $\mb{T}(A)$ occurs in $\mb{T}(A)$ exactly $v^{\br(N)-\br(A)}$ times.
 We also call such $\mb{T}$ a $\voa(M,v)$.
\end{definition}

\begin{remark}
By definition, for a matroid $M=(N,\br)$ with $\mc{C}$ the family of
circuits, a $\voa(M, v)$ is an $\oa(v^{\br(N)}; t,n,v)$, where
\rv{$t=\min_{C\in\mc{C}}|C|-1$} and $n=|N|$. \rv{Note that for such an OA,
  $\lambda=v^{\br(N)-t}$, and thus a $\voa(M, v)$ coincides an
  $\oa(t,n,v)$ if and only if $\br(N)=t$, and therefore $M$ is an $U_{t,n}$.}
On the other hand, two OAs with
the same parameters may be VOAs induced by two distinct matroids as
long as they have the same rank and the same size of the smallest
circuit. This is because an OA only considers a constant strength $t$
for each of its subarray with $t$ columns, while a VOA considers
variable strength $\br(A)$ for any subarray whose columns \rv{are} indexed by
$A$. Thus, a $\voa(M, v)$ can be considered as a refined
characterization of an $\oa(v^{\br(N)}; t,n,v)$.
\end{remark}


\begin{example}
  \label{asdf1}
  For the matroid $M$ in Example \ref{asdf}, we can construct a
  $\voa(M,2)$ below.
 \begin{equation*}
   \begin{matrix}
     0 & 0 & 0 & 0 & 0 & 0 \\
     1 & 0 & 0 & 1 & 0 & 1 \\     
     0 & 1 & 0 & 1 & 0 & 1 \\
     1 & 1 & 0 & 0 & 0 & 0 \\ 
     0 & 0 & 1 & 1 & 0 & 1 \\
     1 & 0 & 1 & 0 & 0 & 0 \\
     0 & 1 & 1 & 0 & 0 & 0 \\
     1 & 1 & 1 & 1 & 0 & 1 \\
     0 & 0 & 0 & 0 & 1 & 1 \\
     1 & 0 & 0 & 1 & 1 & 0 \\
     0 & 1 & 0 & 1 & 1 & 0 \\
     1 & 1 & 0 & 0 & 1 & 1 \\
     0 & 0 & 1 & 1 & 1 & 0 \\
     1 & 0 & 1 & 0 & 1 & 1 \\
     0 & 1 & 1 & 0 & 1 & 1 \\
     1 & 1 & 1 & 1 & 1 & 0
   \end{matrix}
 \end{equation*}
\rv{Note that this $\voa(M,2)$ is also an $\oa(16;2,6,2)$.}
\end{example}

The definition of VOA immediately implies the following lemma.

\begin{lemma}
  \label{lem1}
For a $\voa(M,v)$ $\mb{T}$ with matroid $M=(N,\br)$ and integer
 $v\ge 2$,
  \begin{enumerate}
  \item a subset $I\subseteq N$ is an independent set of $M$ if and only
    if each row of $\mb{T}(I)$
occurs in $\mb{T}(I)$ exactly
$v^{\br(N)-|I|}$ times;\label{ind}
\item a subset $B\subseteq N$ with $|B|=\br(N)$ is a basis of $M$ if and only if each row of
  $\mb{T}(B)$ occurs in $\mb{T}(B)$ exactly once; \label{bas}
  \item a subset $C\subseteq N$ is a circuit $C$ of $M$ if and only if each row of $\mb{T}(C)$
occurs in $\mb{T}(C)$ exactly
$v^{\br(N)-|C|+1}$ times, and for each $e\in C$, each row of
$\mb{T}(C-e)$ occurs in $\mb{T}(C-e)$ exactly
$v^{\br(N)-|C|+1}$ times. \label{cir}
  \end{enumerate}
\end{lemma}


For a random vector $(X_i, i\in N)$, without loss of generality, we
assume each random variable $X_i$ is distributed on \rv{$\SetZ_{v_i}$}.
For each $x\in\SetZ_{v_i}$,  we assume $p_{X_i} (x) > 0$.
By the following theorem in \cite{CCB21}, we characterize matroidal
entropy functions by VOAs.

\begin{theorem}\rv{\cite{CCB21}}
\label{thm} 
  A random vector $\mb{X}=(X_i: i\in N)$ characterizes matroidal entropy
  function $\log v\cdot \br_M$ for a connected matroid $M=(N,\br)$ with rank
  $\mb{r}(N)\ge 2$ if and only if random variable $Y=\mb{X}$ is uniformly distributed
  on the rows of a $\voa(M,v)$. 
\end{theorem}

\begin{remark}
  Theorem \ref{thm} characterizes matroidal entropy functions induced
  by connected matroid $M=(N,\br)$ with rank
  $\mb{r}(N)\ge 2$.
  For a matroid with rank $1$, it is a matroid
  containing $U_{1,n'}$ on $N'\subseteq N$ as a submatroid, where
  $n'=|N|$ and $e\in N\setminus N'$ as loops. They can be
  characterized by $(X_i, i\in N)$, where $X_i=X, i\in N'$, $X$ an
  arbitrary random variable with $H(X)=\log v$,  and $X_i$ a
  constant $i\in N\setminus N'$. For a disconnected matroid, it can
  be written as the directed sum of their connected components and
  its p-characteristic set is the intersection of the p-characteristic
  set of these connected components (See Subsection \ref{12sum} for
  more about direct sum).  
\end{remark}

By Theorem \ref{thm},  whether $\log v\cdot \br_M\in \Gamma^*_n$ or
whether $v\in \chi_M$ for a matroid
$M=(N,\br_M)$ depends on whether $\voa(M,v)$ exists. So in the following sections, we utilize matroid theory and
the constructions of VOAs to determine the p-characteristic set
$\chi_M$ for a matroid $M$.

\rv{As we discussed in the remark after Definition \ref{dvoa},} when the matroid is a uniform matroid $U_{t,n}$, it can be seen that
an array $\mb{T}$ is a $\voa(U_{t,m},v)$ if and only if it is an
$\oa(t,n,v)$. 

\begin{example}
  It is well-known in combinatorial design theory that an $\oa(2,4,v)$ exists if and only if $v\neq 2, 6$.
  The nonexistence of $\oa(2, 4, 6)$ is the famous Euler’s 36 officer
  problem. Thus
  \begin{equation*}
    \chi_{U_{2,4}} =\{v\in \SetZ : v\ge 3, v\neq 6\}.
  \end{equation*}
  The following is an $\oa(2,4,3)$.
  \begin{equation*}
    \begin{matrix}
      0 & 0 & 0 & 0 \\
      0 & 1 & 1 & 1 \\
      0 & 2 & 2 & 2 \\
      1 & 0 & 2 & 1 \\
      1 & 1 & 0 & 2 \\
      1 & 2 & 1 & 0 \\
      2 & 0 & 1 & 2\\
      2 & 1 & 2 & 0\\
      2 & 2 & 0 & 1
    \end{matrix}
  \end{equation*}

  For an $\oa(2,5,v)$, it is well known that it does not exist if
  $v=2,3,6$, it is unknown whether it exists if $v=10$ and it \rv{exists}
  otherwise. So 
 \begin{equation*}
    \{v\in \SetZ : v\ge 4, v\neq 6,10\}\subseteq\chi_{U_{2,5}} \subseteq \{v\in \SetZ : v\ge 4, v\neq 6\}.
  \end{equation*}
  
\end{example}

\section{Constructions}
\label{con}
Constructions of matroids or matroid operations are mappings
from a set of given matroids to a new matroid.
In this section, we discuss some basic matroid constructions and their
corresponding VOA constructions, and leverage them to determine the
p-characteristic set of the constructed matroid. The p-characteristic set
of the matroids by these basic constructions help us build up the tool
box to determine more classes of matroids which will be discussed in
Section \ref{pst}.


\subsection{Deletions, contractions and minors}
\label{dcm}

In this subsection, we introduce the VOA constructions of three
unitary matroid operations, that is, deletion, contraction and their
composition, i.e., taking minor of a matroid.

\begin{definition}[Matroid deletion]
  Given a matroid $M=(N,\mb{r})$ and $S \subseteq N$, the
  matroid $M\setminus S=(N',\mb{r}')$ with $N'=N\setminus S$ and 
  \begin{equation*}
    \mb{r}'(A)=\mb{r}(A),\quad \forall A\subseteq N'
  \end{equation*}
is called the deletion of $S$ from $M$ or the \emph{restriction} of $M$ on
$N'$.
\end{definition}

By definition, each row of $\mb{T}(A)$ occurs $v^{\br(N)-\br(A)}$
times. For any array $\mb{T}'$, we denote $\ded(\mb{T}')$ as the array
whose rows are those of $\mb{T}'$ with each occurring exactly
once, and call it \emph{deduplication} of $\mb{T}'$. 

\begin{construction}[Deletion]
  For a $\voa(M,v)$ $\mb{T}$ and $S\subseteq N$, let
  \begin{equation*}
    \mb{T}\setminus S=\ded(\mb{T}(N')),
  \end{equation*}
   where $N'=N\setminus S$. We call $\mb{T}\setminus S$ the deletion
   of $S$ from $\mb{T}$ or the \emph{restriction} of $\mb{T}$ to
$N'$.
\end{construction}


It is easy to be verified that  $\mb{T}\setminus S$
is a $v^{\br'(N')}\times |N'|$ array and hence the following
proposition.

\begin{proposition}
For a $\voa(M,v)$ $\mb{T}$ and $S\subseteq N$, $\mb{T}\setminus S$
is a $\voa(M\setminus S, v)$.
\end{proposition}

\begin{example}
  \label{asdf5}
  Let $M=(N, \mb{r})$ in Example \ref{asdf}, where
  $N=\{1,2,3,4,5,6\}$. Let $\mb{T}$ be the $\voa(M, 2)$ in
 Example \ref{asdf1}.
 \begin{equation*}
   \begin{matrix}
     {\color{blue} 0}  & {\color{blue} 0}  & {\color{blue} 0}  & {\color{blue} 0}  &  {\color{gray} 0} &  {\color{gray} 0}  \\
     {\color{blue} 1}  & {\color{blue} 0}  & {\color{blue} 0}  &  {\color{blue} 1} &  {\color{gray} 0}  &  {\color{gray} 1} \\     
     {\color{blue} 0}  & {\color{blue} 1}  & {\color{blue} 0}  &  {\color{blue} 1} &  {\color{gray} 0}  &  {\color{gray} 1}  \\
     {\color{blue} 1}  & {\color{blue} 1}  & {\color{blue} 0}  & {\color{blue} 0}  &  {\color{gray} 0}  &  {\color{gray} 0}  \\ 
     {\color{blue} 0}  & {\color{blue} 0}  &  {\color{blue} 1}  &  {\color{blue} 1}  &  {\color{gray} 0}  &  {\color{gray} 1}  \\
     {\color{blue} 1}  & {\color{blue} 0}  &  {\color{blue} 1}  & {\color{blue} 0}  &  {\color{gray} 0}  &  {\color{gray} 0}  \\
     {\color{blue} 0}  & {\color{blue} 1}  &  {\color{blue} 1}  & {\color{blue} 0}  &  {\color{gray} 0}  &  {\color{gray} 0}  \\
     {\color{blue} 1} & {\color{blue} 1} & {\color{blue} 1} & {\color{blue} 1} &  {\color{gray} 0} &  {\color{gray} 1}  \\
     0 & 0 & 0 & 0 &  {\color{gray} 1}  &  {\color{gray} 1}  \\
     1 & 0 & 0 & 1 &  {\color{gray} 1}  & {\color{gray} 0} \\
     0 & 1 & 0 & 1 &  {\color{gray} 1}  & {\color{gray} 0} \\
     1 & 1 & 0 & 0 &  {\color{gray} 1}  & {\color{gray} 1} \\
     0 & 0 & 1 & 1 &  {\color{gray} 1}  & {\color{gray} 0} \\
     1 & 0 & 1 & 0 &  {\color{gray} 1}  & {\color{gray} 1} \\
     0 & 1 & 1 & 0 &  {\color{gray} 1}  & {\color{gray} 1} \\
     1 & 1 & 1 & 1 &  {\color{gray} 1}  & {\color{gray} 0}
   \end{matrix}
 \end{equation*}
 Let $S=\{5,6\}$. Then
  $M_1=M\setminus S$ is a $U_{3,4}$ in $N_1=N\setminus S=\{1,2,3,4\}$.
Deleting the fifth and sixth columns of $\mb{T}$, we obtain an
array $\mb{T}(N_1)$ with each row occurring twice. Keeping one
copy of each row, we obtain $\mb{T}_1=\mb{T}\setminus S$ which is a
$\voa(M_1,2)$ below.
 \begin{equation*}
   \begin{matrix}
     {\color{blue} 0}  & {\color{blue} 0}  & {\color{blue} 0}  & {\color{blue} 0}  \\
     {\color{blue} 1}  & {\color{blue} 0}  & {\color{blue} 0}  &  {\color{blue} 1} \\     
     {\color{blue} 0}  & {\color{blue} 1}  & {\color{blue} 0}  &  {\color{blue} 1} \\
     {\color{blue} 1}  & {\color{blue} 1}  & {\color{blue} 0}  & {\color{blue} 0}   \\ 
     {\color{blue} 0}  & {\color{blue} 0}  &  {\color{blue} 1}  &  {\color{blue} 1}  \\
     {\color{blue} 1}  & {\color{blue} 0}  &  {\color{blue} 1}  & {\color{blue} 0} \\
     {\color{blue} 0}  & {\color{blue} 1}  &  {\color{blue} 1}  & {\color{blue} 0} \\
      {\color{blue} 1} & {\color{blue} 1} & {\color{blue} 1} & {\color{blue} 1} 
   \end{matrix}
 \end{equation*}

\end{example}

\begin{definition}[Matroid contraction]
  Given a matroid $M=(N,\mb{r})$ and $S\subseteq N$, the
  matroid $M/S=(N',\mb{r}')$ with $N'=N\setminus S$ and
  \begin{equation*}
    \mb{r}'(A)=\br(A\cup S)-\br(S),\quad\forall  A\subseteq N'
  \end{equation*}
is called the \emph{contraction} of $S$ from $M$.
\end{definition}

For a row $\mb{b}$ indexed by $N$ and $A\subseteq N$, let $\mb{b}(A)$
denote the sub-row of $\mb{b}$ with entries indexed by $A$.
\begin{construction}[Contraction]
  For a $\voa(M,v)$ $\mb{T}$ and $S\subseteq N$, 
let $\mb{a}$ be a row of $\mb{T}(S)$. We denote by
$\mb{T}_{|S:\mb{a}}$ the array whose rows are $\mb{c}(N\setminus S)$ with
$\mb{c}$ the rows of $\mb{T}$ and $\mb{c}(S)=\mb{a}$. We call
$\mb{T}_{|S:\mb{a}}$ the contraction of $S$ from $\mb{T}$ to $\mb{a}$.
\end{construction}

\begin{proposition}
  \label{pcont}
    For a $\voa(M,v)$ $\mb{T}$ and $S\subseteq N$, $\mb{T}_{|S:\mb{a}}$
is a $\voa(M/ S, v)$ where $\mb{a}$ is any row of $\mb{T}(S)$.
\end{proposition}
\begin{proof}
  By definition, there are $v^{\br(S)}$ different rows in $\mb{T}(S)$,
  each of which occurs $v^{\br(N)-\br(S)}$ times. Consider
  $\mb{T}_{|S:\mb{a}}$. It can be seen it is a $v^{\br(N)-\br(S)}\times
  |N'|$ array, where $N'=N\setminus S$. Now for each $A\subseteq
  N'\subseteq N$, each row of the subarray $\mb{T}(A\cup S)$ occurs exactly
  $v^{\br(N)-\br(A\cup S)}$ times. Among them,
  $v^{\br(N)-\br(S)}$ rows contain $\mb{a}$. Thus, each row of
  $\mb{T}_{|S:\mb{a}}(A)$ occurs exactly
  $v^{\br(N)-\br(A\cup S)}= v^{(\br(N)-\br(S))-(\br(A\cup S)-\br(S))}=v^{\br'(N')-\br'(A)}$
  times, which proves the proposition. 
\end{proof}

\begin{example}
  \label{t2t2}
  Let $M_1=(N_1,\mb{r})$ be a $U_{3,4}$, where
  $N_1=\{1,2,3,4\}$ and $\mb{T}_1$ is the $\voa(M_1,2)$ in Example \ref{asdf5}.
   \begin{equation*}
   \begin{matrix}
     {\color{blue} 0}  & {\color{blue} 0}  & {\color{blue} 0}  & {\color{red} 0}  \\
    1  & 0& 0 &  1\\     
    0  &1  &0  &  1\\
     {\color{blue} 1}  & {\color{blue} 1}  & {\color{blue} 0}  & {\color{red} 0}   \\ 
     0  & { 0}  &  { 1}  &  1  \\
     {\color{blue} 1}  & {\color{blue} 0}  &  {\color{blue} 1}  & {\color{red} 0} \\
     {\color{blue} 0}  & {\color{blue} 1}  &  {\color{blue} 1}  & {\color{red} 0} \\
      { 1} & {1} & { 1} & 1
   \end{matrix}
 \end{equation*}
 Let $S=\{4\}$. Then $M_2=M_1/S$ is a $U_{2,3}$ on
 $N_2=\{1,2,3\}$. It can be checked that $\mb{T}_2=\mb{T}_{1|S:0}$
 is a $\voa(U_{2,3},2)$ below.
   \begin{equation*}
   \begin{matrix}
     {\color{blue} 0}  & {\color{blue} 0}  & {\color{blue} 0}  \\
     {\color{blue} 1}  & {\color{blue} 1}  & {\color{blue} 0}  \\ 
     {\color{blue} 1}  & {\color{blue} 0}  &  {\color{blue} 1} \\
     {\color{blue} 0}  & {\color{blue} 1}  &  {\color{blue} 1} 
   \end{matrix}
 \end{equation*}
 
\end{example}

\noindent {\bf  Remark} For a matroidal entropy function $\log v\cdot \br_M$ with
characterizing random vector $\mb{X}=(X_i:i\in N)$, its contraction
$\log v
\cdot \br_{M/S}$ is the entropy function of $\mb{Y}=(Y_i: i\in N\setminus
S)$, where $Y_A=X_A|X_S$ for any $A\subset N\setminus S$, that is,
each entry of $\log v
\cdot \br_{M/S}$ is equal to the conditional entropy $H(X_A|X_S)$. Thus, if
$\mb{X}$ is distributed on the rows of a $\voa(M,v)$ $\mb{T}$, then
$\mb{Y}$ is distributed on the rows of a $\voa(M/S,v)$
$\mb{T}|_{S:\mb{a}}$ for any $\mb{a}$ in $\mb{T}(S)$. 

Deletion and contraction are two ways of taking substructure of a
matroid. These two operations can be \rvv{composed}, resulting a general
substructure ``minor'' of a
matroid.
According to \cite[Proposition 3.1.25(iii)]{O11}, for a sequence of disjoint
$S_1, S_2,\ldots, S_k\subseteq N$, $M$ being deleted or contracted by $S_i$, the
result can be written in the form of $M\setminus S/T$, where $S$ is the
union of the deleted $S_i$ and $T$ is the union of the contracted
$S_j$. Such $M\setminus S/T$ is called a \emph{minor} of
$M$. Similarly, we can construct any minor of a VOA.

\begin{construction}[Minor]
  Let $\mb{T}$ be a $\voa(M,v)$. For any disjoint $S, T\subseteq N$, and
  $\mb{a}=\mb{c}(T)$ for some row $\mb{c}$ of $\mb{T}$,
  $\mb{T}\setminus S|_{T:\mb{a}}$ is a a minor of $\mb{T}$.
\end{construction}

\begin{example}
  The $\voa(U_{3,4},2)$ $\mb{T}_1$ in Example \ref{asdf5} and $\voa(U_{2,3},2)$ $\mb{T}_2$ in
  Example \ref{t2t2} are both minors of the $\voa(M,2)$ in Example \ref{asdf1}. 
\end{example}

\begin{theorem}
  \label{minor}
  Let $M$ be a matroid and $M'$ be its minor. Then $\chi_{M}\subseteq \chi_{M'}$.
\end{theorem}
\begin{proof}
  As discussed above, $M'=M\setminus S/T$ for some disjoint $S,
  T\subseteq N$. Hence, if there exists a $\voa(M,v)$ $\mb{T}$,
  we can construct a $\voa(M',v)$ $\mb{T}\setminus S|_{T:\mb{a}}$ for some
  $\mb{a}=\mb{b}(T)$, where $\mb{b}$ is a row of $\mb{T}$.
  It follows that $\chi_{M}\subseteq \chi_{M'}$.
\end{proof}

\begin{remark}
  For a class $\mc{M}$ of matroids, excluded minors \rv{(or forbidden
  minors in some literatures)}
  of $\mc{M}$ take an important role to characterize it. See more
  discussion about this topic in Subsection \ref{extrac}.
\end{remark}





\subsection{Series and parallel connections}
\label{bop}

In this subsection, we
consider the VOAs of another two matroid constructions, \rv{namely},
series connection, parallel connection\rv{,} which are binary operations on
matroids. \rv{In the following of this paper, for a matroid $M$, an
  array $\mb{T}$ and a set $S$, we write $M\setminus S$,
    $\mb{T}\setminus $, $M/ S$ and $\mb{T}|_{S:\mb{a}}$ as $M\setminus
    p$, $\mb{T}\setminus p$, $M/p$ and $\mb{T}_{p:\mb{a}}$,
    respectively for simplicity when $S=\{p\}$ is singleton, }

\begin{definition}[Series and parallel connections of matroids]
  For two matroids $M_1=(N_1, \br_1)$ and $M_2=(N_2, \br_2)$ with $p_i\in
  N_i$, $p_i$ neither loops nor coloops, $i=1,2$,  and any
  $p\not\in N_1\cup N_2$ the
  \emph{series connection} $S((M_1;p_1),(M_2;p_2))$ of
  $M_1$ and $M_2$ with respect to \emph{base points} $p_1$ and $p_2$
  is a matroid with ground set $N\triangleq (N_1\setminus p_1)\cup
  (N_2\setminus p_2)\cup p$ and family of circuits
  \begin{align}
    \label{eq:1}
    \mc{C}_S=  &\mc{C}(M_1\setminus p_1)\cup\mc{C}(M_2\setminus p_2)\nonumber\\
       &\cup \{(C_1-p_1)\cup(C_2-p_2)\cup p: C_i\in
                \mc{C}(M_i), i=1,2\}
  \end{align}
and  the
  \emph{parallel connection} $P((M_1;p_1),(M_2;p_2))$ of
  $M_1$ and $M_2$ with respect to \emph{base points} $p_1$ and $p_2$
  is a matroid with ground set $N$ and family of circuits
  \begin{align}
    \label{eq:2}
    \mc{C}_P= & \mc{C}(M_1\setminus p_1)\cup\mc{C}(M_2\setminus p_2)
       \cup \{(C_1-p_1)\cup p: C_1\in
                \mc{C}(M_1)\}\nonumber\\
    &          \cup \{(C_2-p_2)\cup p: C_2\in\mc{C}(M_2).\}
  \end{align}

\end{definition}

The validity that $\mc{C}_S$ and $\mc{C}_P$ are families of circuits
of matroids is due to \cite[Proposition 7.1.4]{O11}. We write
$S(M_1,M_2)$ for $S((M_1;p_1),(M_2;p_2))$ and $P(M_1,M_2)$ for
$P((M_1;p_1),(M_2;p_2))$ if there is no ambiguity. By
\cite[Proposition 7.1.15]{O11}, the rank of $S(M_1,M_2)$ is
\begin{equation*}
  r_S=\br_1(N_1)+\br_2(N_2)
\end{equation*}
and the rank of $P(M_1,M_2)$ is
\begin{equation*}
  r_{P}=\br_1(N_1)+\br_2(N_2)-1.
\end{equation*}

\begin{construction}[Series connection]
  Let $\mb{T}_1$ be a $\voa(M_1, v)$ and $\mb{T}_2$ be a $\voa(M_2, v)$,
where matroid $M_1=(N_1,\br_1)$
and $M_2=(N_2,\br_2)$ with $p_i\in N_i,i=1,2$ and $v$ is an integer.
Let $\mb{U}$ be any $\oa(2,3,v)$.
We construct an
$v^{r_S}\times (|N_1|+|N_2|-1)$ array $\mb{T}$ with columns
indexed by $N=(N_1\setminus p_1)\cup
(N_2\setminus p_2)\cup p$ according to the following rule.
\begin{itemize}
\item For any row $\mb{a}_1$ of $\mb{T}_1$
and $\mb{a}_2$ of $\mb{T}_2$, we construct a row $\mb{b}$ of $\mb{T}$
such that $\mb{b}(N_1\setminus p_1)=\mb{a}_1 (N_1\setminus p_1)$,
$\mb{b}(N_2\setminus p_2)=\mb{a}_2 (N_2\setminus p_2)$ and
$(\mb{a}_1(p_1), \mb{a}_2(p_2), \mb{b}(p))$ forms a row of $\mb{U}$.
\end{itemize}
We denote such constructed $\mb{T}$ by
$S((\mb{T}_1;p_1),(\mb{T}_2;p_2))$ or $S(\mb{T}_1,\mb{T}_2)$ if there
is no ambiguity.  It can be checked that $\mb{T}$ is a VOA.
\end{construction}

\begin{proposition}
  \label{psc}
 For a $\voa(M_1, v)$ $\mb{T}_1$ and a $\voa(M_2, v)$ $\mb{T}_2$, the array $S((\mb{T}_1;p_1),(\mb{T}_2;p_2))$ is a $\voa(S((M_1;p_1),(M_2,p_2)),v)$.
\end{proposition}
\begin{proof}
  To prove the proposition, for $S((\mb{T}_1;p_1),(\mb{T}_2;p_2))$,
  denoted by
  $\mb{T}$ for short in the proof, we only need to check that for each circuit $C$ of
  $S((M_1;p_1),(M_2,p_2))$ (cf. \eqref{eq:1}), $\mb{T}(C)$ satisfied the
  if part of Lemma \ref{lem1}(\ref{cir}). If so, 
  we say $\mb{T}$ represents $C$ in $S(M_1,M_2)$. 
  
We first consider $C\in\mc{C}(M_1\setminus p_1)$. As $C$ is a circuit
of $M_1\setminus p_1$, $C$ is also a circuit of $M_1$.
By the only if
part of Lemma \ref{lem1}(\ref{cir}), each row of $\mb{T}_1(C)$ occurs exactly
$v^{\br_1(N_1)-|C|+1}$ times in $\mb{T}_1(C)$. According to the
construction of $\mb{T}$, each row of $\mb{T}_1(C)$ is duplicated
$v^{\br_2(N_2)}$ times in $\mb{T}$. Thus each row of $\mb{T}(C)$ occurs
exactly $v^{r_S-|C|+1}$ times. By the same argument, for any $e\in
C$,  each row of $\mb{T}(C-e)$ occurs
exactly $v^{r_S-|C|+1}$ times. Thus $\mb{T}$ represents $C$ in $S(M_1,M_2)$. 
By symmetry, we can prove that $\mb{T}$ represents
$C\in\mc{C}(M_2\setminus p_2)$ in $S(M_1,M_2)$.

Now we consider $C=(C_1-p_1)\cup (C_2-p_2)\cup p$ for $C_i\in
                \mc{C}(M_i), i=1,2$. As $C_i-p_i$ is an independent
                set of $M_i$, by Lemma \ref{lem1}(\ref{ind}), each row $\mb{a}_i$
                of $\mb{T}_i(C_i-p_i)$ occurs $v^{\br_i(N_i)-|C_i|+1}$
                times. According to the construction of $\mb{T}$,
                each row $\mb{a}$ in $\mb{T}((C_1-p_1)\cup (C_2-p_2))$
                is the concatenation of some row $\mb{a}_1$ in
                $\mb{T}_1(C_1-p_1)$ and some row $\mb{a}_2$
                in $\mb{T}_2(C_2-p_2)$, thus $\mb{a}$ occurs exactly
                $v^{r_S-|C_1|-|C_2|+2}$ times. For a row  $\mb{b}_i$
                in $\mb{T}_i(C_i)$, $\mb{b}_i(p_i)$ is uniquely
                determined by $\mb{b}_i(C_i-p_i)$.
                Now for a row $\mb{b}$ in
                $\mb{T}((C_1-p_1)\cup (C_2-p_2)\cup p)$ with
                $\mb{b}(C_i-p_i)=\mb{b}_i(C_i-p_i)$,
                as
                $\mb{b}_1(p_1), \mb{b}_2(p_2), \mb{b}(p)$ forms a row
                in $\mb{U}$, $\mb{b}(p)$ is uniquely determined by
                $\mb{b}_1(p_1)$ and $\mb{b}_2(p_2)$, and thus uniquely
                determined by $\mb{b}(C_i-p_i), i=1,2$. So  $\mb{b}$
                occurs in
                $\mb{T}((C_1-p_1)\cup (C_2-p_2)\cup p)$ the same
                number of times as
                $\mb{b}((C_1-p_1)\cup (C_2-p_2))$ occurs in
                $\mb{T}((C_1-p_1)\cup (C_2-p_2))$, i.e.,
                $v^{r_S-|C_1|-|C_2|+2}$ times. 

                To prove the
                proposition, we now only need to check that
                for any $e\in C_i-p_i$, say $e\in C_1-p_1$, any row $\mb{a}$ in
                $\mb{T}'=\mb{T}((C_1-p_1-e)\cup (C_2-p_2)\cup p)$ also occurs
                exactly $v^{r_S-|C_1|-|C_2|+2}$ times.
               Let $\mb{T}''=\mb{T}_1(C_1-e)\oplus \mb{T}_2(C_2-p_2)$. (See the definition of direct sum
                ``$\oplus$'' of two VOAs in the next subsection. Here we abuse the notation 
                 for two arrays.) 
                As $C_1-e$
                is independent in $M_1$ and $C_2-p_2$ is independent
                in $M_2$, each row of $\mb{T}''$ occurs
                $v^{r_S-|C_1|-|C_2|+2}$ times.
                Consider  row $\ba$ of
                $\mb{T}''$,  row $\bb$ of $\mb{T}'$ with
                $\bb(C_1-e-p_1)=\ba(C_1-e-p_1)$ and
                $\bb(C_2-p_2)=\ba(C_2-p_2)$ and row $\bc$ of
                $\mb{T}_2$ with $\bc(C_2-p_2)=\ba(C_2-p_2)$. As
                $\bc(p_2)$ is uniquely determined by $\bc(C_2-p_2)$ or
                $\ba(C_2-p_2)$, and $\bb(p)$ is uniquely determined by
                $\ba(p_1)$ and $\bc(p_2)$, $\bb$ is uniquely
                determined by $\ba$. Replacing each $\ba$ by $\bb$, we
                obtain that each row of $\mb{T}'$ occurs
                exactly $v^{r_S-|C_1|-|C_2|+2}$ times.                
                Thus we prove $\mb{T}$
                represents $C=(C_1-p_1)\cup (C_2-p_2)\cup p$ in $S(M_1,M_2)$ and so
                the proposition is proved.
              \end{proof}

\begin{example}
  Let $M_1$ be a $U_{3,4}$ on $N_1=\{1,2,3,4\}$ and $M_2$ be a
  $U_{2,3}$ on $N_2=\{4,5,6\}$. It can be seen that
  $S((M_1;4),(M_2;4))$ is a $U_{5,6}$.
Let $\mb{T}_1$ be a $\voa(M_1,2)$ 
  below
   \begin{equation*}
     \begin{matrix}
 {\color{blue} 0}  & {\color{blue} 0}  & {\color{blue} 0}
            & \overline{\color{blue} 0}\\

      {\color{green} 1} &  {\color{green} 0}  &  {\color{green} 0} & \overline{{\color{green} 1}} \\     
     0 & 1 & 0 & 1 \\
     1 & 1 & 0 & 0 \\ 
     0 & 0 & 1 & 1 \\
     1 & 0 & 1 & 0 \\
     0 & 1 & 1 & 0 \\
     1 & 1 & 1 & 1 
   \end{matrix}
 \end{equation*}
  and $\mb{T}_2$ be a $\voa(M_2,2)$ below
   \begin{equation*}
   \begin{matrix}
    \hat{\color{red} 0}  & {\color{red} 0}  & {\color{red} 0}  \\
     \hat{\color{darkred} 1} &  {\color{darkred} 1} &  {\color{darkred} 0}  \\
     1 &  0 & 1 \\ 
     0  & 1 & 1
   \end{matrix}
 \end{equation*}
Let $U$ be a $\voa(U_{2,3},2)$
   \begin{equation*}
   \begin{matrix}
     \overline{\color{blue} 0} & \hat{\color{red} 0} & \boxed{\color{purple} 0} \\
      \overline{\color{blue} 0} &\hat{\color{darkred} 1}  & \boxed{\color{purple} 1}\\
     \overline{{\color{green} 1}} &  \hat{\color{red} 0} &   \boxed{\color{orange} 1} \\ 
     \overline{{\color{green} 1}} & \hat{\color{darkred} 1}& \boxed{\color{orange} 0} 
   \end{matrix}
 \end{equation*}
 Then a $2^5\times 6$  array $\mb{T}=S((\mb{T}_1,4);(\mb{T}_2,4))$ constructed below is an
 $\oa(5,6,2)=\voa(S((M_1,4);(M_2),4))$.
   \begin{equation*}
     \begin{matrix}
          {\color{blue} 0}  & {\color{blue} 0}  & {\color{blue} 0}  &
          \boxed{\color{purple} 0}   & {\color{red} 0}  &
          {\color{red} 0} \\

                    {\color{blue} 0}  & {\color{blue} 0}  & {\color{blue} 0}  &
          \boxed{\color{purple} 1}   & {\color{darkred} 1}  &
          {\color{darkred} 0} \\
          0 & 0 & 0 & 1 & 0 & 1 \\
          0 & 0 & 0 & 0 & 1 & 1 \\
     {\color{green} 1} &  {\color{green} 0}  &  {\color{green} 0} &
     \boxed{\color{orange} 1}   & {\color{red} 0}  & {\color{red} 0}
     \\
          {\color{green} 1} &  {\color{green} 0}  &  {\color{green} 0} &
     \boxed{\color{orange} 0}   & {\color{darkred} 1}  & {\color{darkred} 0}
     \\
     1 & 0 & 0 & 0 &  0 & 1 \\
      1 & 0 & 0 & 1 &  1 & 1 \\     
      &  &  \vdots &  &   & \\
     1 & 1 & 1 & 1 & 1 &1 
   \end{matrix}
 \end{equation*}
To illustrate the construction of $\mb{T}$, we use those row of
$\mb{T}$ constructed by the first two rows of $\mb{T}_1$ (blue and
green, resp.) and first rows of $\mb{T}_2$ (red and dark red,
resp.). For the first row of $\mb{T}_1$, i.e.,
``${\color{blue} 0}  \quad  {\color{blue} 0}  \quad   {\color{blue} 0}
            \quad  \overline{\color{blue} 0}$'', and for the first row
            of $\mb{T}_2$, i.e., ``$\hat{\color{red} 0}  \quad
            {\color{red} 0}  \quad {\color{red} 0}$'', the entries
            indexed by $4$ are ``$ \overline{\color{blue} 0} $'' and
            ``$\hat{\color{red} 0}$ '', respectively. We check that
            the row in $U$ with first two entries ``$ \overline{\color{blue} 0} $'' and
            ``$\hat{\color{red} 0}$ '' is its first row, and its third
            entry is ``$\boxed{\color{purple} 0} $''. Therefore, we
            construct the first row of $\mb{T}$, i.e., ``$ {\color{blue} 0}  \quad {\color{blue} 0}  \quad {\color{blue} 0}  \quad
          \boxed{\color{purple} 0}   \quad {\color{red} 0}  \quad
          {\color{red} 0} $''. Similarly, we construct the second row
          of $\mb{T}$ by the first row of $\mb{T}_1$ and the second row of
          $\mb{T}_2$, the fifth row of $\mb{T}$ by the second row of
          $\mb{T}_1$ and the first row of $\mb{T}_2$, the sixth row of
          $\mb{T}$ by the second row of $\mb{T}_1$ and second row of
          $\mb{T}_2$. Other rows of $\mb{T}$ can also be constructed
          by the same method by one row of $\mb{T}_1$ and one row of $\mb{T}_2$.
 
\end{example}

\begin{construction}[Parallel connection]
                Let $\mb{T}_1$ be a $\voa(M_1, v)$ and $\mb{T}_2$ be a $\voa(M_2, v)$,
where matroid $M_1=(N_1,\br_1)$
and $M_2=(N_2,\br_2)$ with $p_i\in N_i,i=1,2$ and $v$ is an integer.
We construct a
$v^{r_P}\times (|N_1|+|N_2|-1)$ array $\mb{T}$ with columns
indexed by $N=(N_1\setminus p_1)\cup (N_2\setminus p_2)\cup p$
according to the following rule.
\begin{itemize}
\item For any row $\mb{a}_1$ of $\mb{T}_1$
and $\mb{a}_2$ of $\mb{T}_2$ with $\mb{a}_1(p_1)=\mb{a}_2(p_2)$, we
construct row $\mb{b}$ of $\mb{T}$
such that $\mb{b}(N_i\setminus p_i)=\mb{a}_i$, $i=1,2$, and $\mb{b}(p)=\mb{a}_1(p_1)$.
\end{itemize}
We denote such constructed $\mb{T}$ by
$P((\mb{T}_1;p_1),(\mb{T}_2;p_2))$ or $P(\mb{T}_1,\mb{T}_2)$ if there
is no ambiguity. It can be checked that $\mb{T}$ is a VOA.
              \end{construction}

\begin{proposition}
    \label{ppc}
 For a $\voa(M_1, v)$ $\mb{T}_1$ and a $\voa(M_2, v)$ $\mb{T}_2$, the
 array $P((\mb{T}_1;p_1),(\mb{T}_2;p_2))$ is a $\voa(P((M_1;p_1),(M_2;p_2)),v)$.
\end{proposition}
\begin{proof}
    To prove the proposition, for $P((\mb{T}_1;p_1),(\mb{T}_2;p_2))$,
  denoted by
  $\mb{T}$ for short in the proof, we only need to check that for each circuit $C$ of
  $P((M_1;p_1),(M_2,p_2))$(cf. \eqref{eq:2}), $\mb{T}(C)$ satisfies the
  if part of Lemma \ref{lem1}(\ref{cir}). If it is so, 
  we say $\mb{T}$ represents $C$ in $P(M_1,M_2)$.

We first consider $C\in \mc{C}(M_1\setminus p_1)$. As $C$ is a circuit
of $M_1\setminus p_1$, $C$ is also a circuit of $M_1$.
By the only if part of Lemma \ref{lem1}(\ref{cir}), each row $\mb{a}$ of $\mb{T}_1(C)$ occurs exactly
$v^{\br_1(N_1)-|C|+1}$ times in $\mb{T}_1(C)$. According to the
construction of $\mb{T}$, each row of $\mb{T}_1$,  as well 
as each row of $\mb{T}_1(C)$ is duplicated
$v^{\br_2(N_2)-1}$ times in $\mb{T}$. Thus, each row of $\mb{T}(C)$
occurs exactly $v^{\br_1(N_1)+\br_2(N_2)-|C|}=v^{r_P-|C|+1}$ times in $\mb{T}(C)$. Now for any
$e\in C$, by the only if part of Lemma \ref{lem1}(\ref{cir}), each row $\mb{a}$ of $\mb{T}_1(C-e)$ occurs exactly
$v^{r_1(N_1)-|C|+1}$ times in $\mb{T}_1(C-e)$. By the same argument
above, each row of $\mb{T}(C-e)$
occurs exactly $v^{\br_P-|C|+1}$ times in $\mb{T}(C-e)$. Thus $\mb{T}$
represents $C$ in $P(M_1,M_2)$. By symmetry $\mb{T}$
represents $C\in \mc{C}(M_2\setminus p_2)$ in $P(M_1,M_2)$.

Now we consider $C=(C_1-p_1)\cup p, p_1\in C_1\in\mc{C}(M_1)$. As $C_1$
is a circuit of $M_1$, each row $\mb{a}_1$ of $\mb{T}_1(C_1)$ occurs in
$\mb{T}_1(C_1)$ exactly
$v^{\br_1(N_1)-|C|+1}$ times. According to the construction of $\mb{T}$, for a row
$\mb{a}$ in $\mb{T}$ with $\mb{b}(C-p)=\mb{a}(C_1-p_1)$,
$\mb{a}(p)= \mb{a}_1(p_1)$. Moreover, as $p_2$ is not a loop of
$\mb{T}$, there exists $v^{\br_2(N_2)-1}$ rows $\mb{a}_2$ with
$\mb{a}_2(p_2)=\mb{a}_1(p_1)$. Thus, $\mb{a}_1$ is duplicated
$v^{\br_2(N_2)-1}$ times in $\mb{T}$. So is $\mb{a}_1(C_1)$ or
$\mb{a}(C)$. Hence $\mb{a}(C)$ occurs in $\mb{T}$ exactly
$v^{\br_1(N_1)+\br_2(N_2)-|C|}=v^{r_P-|C|+1}$ times. For any $e\in C$,
either $e\in C-p$ or $e=p$, with the same argument above, we have that
each row of
$\mb{T}(C-e)$ occurs exactly $v^{r_P-|C|+1}$ times. Thus $\mb{T}$
represents $C$ in $P(M_1,M_2)$. By symmetry, $\mb{T}$
represents $C=(C_2-p_2)\cup p, p_2\in C_2\in\mc{C}(M_2)$ in
$P(M_1,M_2)$ which implies the proposition.
\end{proof}

\begin{example}
  Let $M_1$ be a $U_{3,4}$ on $N_1=\{1,2,3,4\}$ and $M_2$ be a
  $U_{2,3}$ on $N_2=\{4,5,6\}$. It can be seen that
  $P((M_1;4),(M_2;4))$ is
the matroid $M$ in Example \ref{asdf}.
Let $\mb{T}_1$ be a $\voa(M_1,2)$ 
below
   \begin{equation*}
     \begin{matrix}
 {\color{blue} 0}  & {\color{blue} 0}  & {\color{blue} 0}
            & \boxed{\color{blue} 0}\\

      {\color{green} 1} &  {\color{green} 0}  &  {\color{green} 0} & \boxed{{\color{green} 1}} \\     
     0 & 1 & 0 & 1 \\
     1 & 1 & 0 & 0 \\ 
     0 & 0 & 1 & 1 \\
     1 & 0 & 1 & 0 \\
     0 & 1 & 1 & 0 \\
     1 & 1 & 1 & 1 
   \end{matrix}
 \end{equation*}

  and $\mb{T}_2$ be a $\voa(M_2,2)$ below
 \begin{equation*}
   \begin{matrix}
    \boxed{\color{red} 0}  & {\color{red} 0}  & {\color{red} 0}  \\
     \boxed{\color{darkred} 1} &  {\color{darkred} 1} &  {\color{darkred} 0}  \\
     1 &  0 & 1 \\ 
     0  & 1 & 1
   \end{matrix}
 \end{equation*}

Then we obtain a $2^4\times 6$ array
$\mb{T}=P((\mb{T}_1,4);(\mb{T}_2,4))$ below, which is a
$\voa(P((M_1;4),(M_2;4)),v)$.
 \begin{equation*}
     \begin{matrix}
          {\color{blue} 0}  & {\color{blue} 0}  & {\color{blue} 0}  &
          \boxed{\color{purple} 0}   & {\color{red} 0}  &
          {\color{red} 0} \\

          0 & 0 & 0 & 0 &1 &1 \\

          {\color{green} 1} &  {\color{green} 0}  &  {\color{green} 0} &
     \boxed{\color{orange} 1}   & {\color{darkred} 1}  & {\color{darkred} 0}
     \\
      1 & 0 & 0 & 1 &  0 & 1 \\     
      &  &  \vdots &  &   & \\
     1 & 1 & 1 & 1 & 1 &1 
   \end{matrix}
 \end{equation*}

\end{example}

\begin{theorem}
  \label{tspc}
  For any matroids $M_1$ and $M_2$,
  $\chi_{S(M_1,M_2)}=\chi_{P(M_1,M_2)}=\chi_{M_1}\cap \chi_{M_2}$.
\end{theorem}
\begin{proof}
  By Propositions \ref{psc} and \ref{ppc}, we see that
  $ \chi_{M_1}\cap \chi_{M_2}\subseteq\chi_{S(M_1,M_2)}$ and
  $\chi_{M_1}\cap \chi_{M_2}\subseteq \chi_{P(M_1,M_2)}$. As
  $M_i=S(M_1,M_2)/ (N_j-p)=P(M_1,M_2)\setminus (N_j-p)$ for $i,j=1,2,
  i\neq j$ \cite[Proposition 7.1.15(ii)]{O11}, both $M_i$ are minors of $P(M_1,M_2)$ and of
  $S(M_1,M_2)$. Thus by Theorem \ref{minor},  $\chi_{S(M_1,M_2)}\chi_{M_1}\subseteq \cap
  \chi_{M_2}$ and $\chi_{P(M_1,M_2)}\subseteq \chi_{M_1}\cap
  \chi_{M_2}$ which implies the theorem.
\end{proof}

\subsection{Direct sum and $2$-sum}
\label{12sum}

In this subsection, we consider two more binary matroid operations direct sum and $2$-sum,  which are
related to the connectivity and higher connectivity of a matroid.

\begin{definition}[Direct sum]
  Let $M_1=(N_1, \br_1)$ and $M_2=(N_2, \br_2)$ be two matroids and
  $N_1$ and $N_2$ disjoint. The matroid $M=(N, \br)$, where $N=N_1\cup
  N_2$ and 
  \begin{equation*}
    \br(A)=\br(A\cap N_1)+\br(A\cap N_2) \quad \text{ for all }
    A\subseteq N
  \end{equation*}
is called \emph{direct sum} or \emph{$1$-sum}  of $M_1$ and $M_2$, and
is denoted by $M_1\oplus M_2$.
\end{definition}

The fact that $\br_M$ defined above forms the rank function of a
matroid is due to \cite[4.2.13]{O11}. By definition, the rank of
$M_1\oplus M_2$ is $\br(N_1)+\br(N_2)$, and the restriction of
$M_1\oplus M_2$ on $N_i$ is $M_i$, $i=1,2$.

\begin{construction}[Direct sum]
  Let $\mb{T}_1$ and $\mb{T}_2$ be $\voa(M_1,v)$ and $\voa(M_2,v)$, 
  respectively. Let $\mb{T}$ be a $v^{\br(N_1)+\br(N_2)}\times
  (|N_1|+|N_2|)$ \rv{array}. For each row $\ba_1$ of $\mb{T}_1$ and $\ba_2$ of
  $\mb{T}_2$, we construct a row $\ba$ of $\mb{T}$ such that
  $\ba(N_1)=\ba_1$ and $\ba(N_2)=\ba_2$. We call such constructed
  $\mb{T}$ \emph{direct sum} or \emph{$1$-sum} of $\mb{T}_1$ and
  $\mb{T}_2$, and denote it by $\mb{T}_1\oplus\mb{T}_2$.
\end{construction}

\begin{proposition}
\label{pdsum}
  For a $\voa(M_1, v)$ $\mb{T}_1$ and a $\voa(M_2, v)$ $\mb{T}_2$, the
  array $\mb{T}=\mb{T}_1\oplus\mb{T}_2$ is a $\voa(M_1\oplus M_2,v)$. 
\end{proposition}
\begin{proof}
  Let $A\subseteq N=N_1\cup N_2$. By definition, $\br(A)=\br_1(A\cap
  N_1)+\br_2(A\cap N_2)$. Consider a row $\ba$ of $\mb{T}(A)$. As
  $\ba(A\cap N_i)$ occurs in $\mb{T}_i(A\cap N_i)$ $v^{\br_i(N)-\br_i(A\cap
    N_i)}$ times, $i=1,2$, $\ba$ occurs in  $\mb{T}(A)$
  $v^{\br(N)-(\br_1(A\cap N_1)+\br_2(A\cap N_2))}$ times, which implies that
  $\mb{T}$ is a $\voa(M_1\oplus M_2,v)$.
\end{proof}

  \begin{theorem}
  \label{tdsum}
  For any matroids $M_1$ and $M_2$, $\chi_{M_1\oplus  M_2}=\chi_{M_1}\cap \chi_{M_2}$.
\end{theorem}
\begin{proof}
  By Proposition \ref{pdsum}, $ \chi_{M_1}\cap \chi_{M_2}\subseteq\chi_{M_1\oplus
    M_2}$. On the other hand, it is
  readily seen that both $M_1$ and $M_2$ are minors of $M_1\oplus
  M_2$. By Theorem \ref{minor}, $\chi_{M_1\oplus M_2}\subseteq
  \chi_{M_i}, i=1,2$ or $\chi_{M_1\oplus M_2}\subseteq
  \chi_{M_1}\cap\chi_{M_2}$. It follows the theorem.
\end{proof}

By \cite[Corollary 4.2.9]{O11}, any matroid can be written as the direct sum of
its connected components. Together with  Theorem \ref{tdsum}, to
determine the $p$-characteristic set of matroids, we only need to
consider connected matroids as building blocks. However, in the
following, we will see that these building blocks can be decomposed
into smaller blocks and each of them can be written as the 2-sum of
these smaller building blocks, i.e., $3$-connected matroids.

\begin{definition}
  For matroids $M_1=(N_1,\br_1)$ and $M_2=(N_2,\br_2)$, the
  \emph{2-sum} of them $M_1\oplus_2M_2$ is defined by $S(M_1,M_2)/p$
  or equivalently $P(M_1,M_2)\setminus p$.
\end{definition}

\begin{construction}[2-sum]
  Let $\mb{T}_1$ be a $\voa(M_1, v)$ and $\mb{T}_2$ be a $\voa(M_2, v)$,
where matroid $M_1=(N_1,\br_1)$
and $M_2=(N_2,\br_2)$ with $p_i\in N_i,i=1,2$ and $v$ is an
integer. We construct $\mb{T}_1\oplus_2 \mb{T}_2$ by
$S(\mb{T}_1,\mb{T}_2) |_{p:a}$ for some $a\in \SetZ_v$, or equivalently
$P(\mb{T}_1,\mb{T}_2)\setminus p$. 
\end{construction}

Note that though $\mb{T}_1\oplus_2
\mb{T}_2$ can be obtained from two different compositions of matroid
operations, they are essentially the same.
We immediately obtain the following
proposition.

\begin{proposition}
  \label{p2sum}
    For a $\voa(M_1, v)$ $\mb{T}_1$ and a $\voa(M_2, v)$ $\mb{T}_2$, $\mb{T}_1\oplus_2 \mb{T}_2$ is a $\voa(M_1\oplus_2 M_2, v)$.
\end{proposition}

\begin{theorem}
  \label{t2sum}
  For any matroids $M_1$ and $M_2$, $\chi_{M_1\oplus_2 M_2}=\chi_{M_1}\cap \chi_{M_2}$.
\end{theorem}
\begin{proof}
  By Proposition \ref{p2sum}, $\chi_{M_1}\cap \chi_{M_2}\subseteq \chi_{M_1\oplus_2
    M_2}$. On the other hand, it is
  readily seen that both $M_1$ and $M_2$ are minors of $M_1\oplus_2
  M_2$. By Theorem \ref{minor}, $\chi_{M_1\oplus_2 M_2}\subseteq
  \chi_{M_i}, i=1,2$. It follows the theorem.
\end{proof}

We consider only connected matroid in \cite{CCB21} because each
matroid can be decomposed into connected matroids and thus its
p-characteristic set is the intersection of its connected components. 
By \cite[Theorem 8.3.10]{O11}, each (2-)connected matroid can be tree
decomposed into 3-connected matroids, circuits and co-circuits. These
3-connected components, circuits and co-circuits are $2$-connected by
$2$-sums. As
circuits and co-circuits can be considered as uniform matroids
$U_{n-1,n}$ and $U_{1,n}$, respectively, whose p-characteristic set is
$\{v\in \SetZ: v\ge 2\}$ as we discussed in \cite{CCB21}, by Theorem
\ref{t2sum}, we
immediately have the following corollary.
\begin{corollary}
  \label{3con}
  The p-characteristic set of a connected matroid is the intersection
  of the p-characteristic set of its $3$-connected components. 
\end{corollary}
By Corollary \ref{3con}, to characterize all matroidal entropy
functions, it is sufficient for us to focus on $3$-connected matroids,
which can be considered as the smallest building blocks.

\section{P-characteristic set of two classes of matroids}
\label{pst}
In this section, we consider two classes of matroids, that is, regular
matroids, whose p-characteristic set is $\{v\in \SetZ: v\ge 2\}$, and
the matroids whose p-characteristic set is the same as $U_{2,4}$,
i.e.,  $\{v\in \SetZ: v\ge 3, v\neq 6\}$.

\subsection{Regular matroid}
\label{srm}
\begin{definition}
  A matroid $M$ is regular if it is represented by a totally
  unimodular matrix, i.e., a matrix over $\mathbb{R}$ for which every
  square submatrix has determinant in $\{-1, 1, 0\}$.
\end{definition}

It has been proved in \cite[Theorem 6.6.3]{O11} that any regular matroid is
representable over any field $\mathbb{F}$. In this subsection, 
we extend this result to partition representability with any degree
$v\ge 2$. 

\begin{lemma}
  \label{lr}
  For an $r\times r$ matrix $\mb{M}$ over the ring $\SetZ_v$,
  $\mb{x}\mapsto \mb{x} \mb{M}$
   maps $\SetZ_v^r$ to  $\SetZ_v^r$ bijectively if and only if the 
  determinant of $\mb{M}$ is a unit in the ring. 
\end{lemma}
\begin{proof}
  By \cite[Corollary 2.21]{B93}, $\mb{M}$ is invertible if and only if its
  determinant is a unit of $\SetZ_v$. As it is easy to be seen that $\mb{x}\mapsto \mb{x} \mb{M}$
   maps $\SetZ_v^r$ to  $\SetZ_v^r$ bijectively if and only if $\mb{M}$ is 
   invertible, the lemma follows. 
\end{proof}


\begin{lemma}
\label{voazv}
Let $M=(N, \br)$ be a regular matroid and $\mb{U}$ its totally unimodular
representation over ring $\SetZ_v$. Then $\mb{x}\mapsto \mb{x}\mb{U}$ maps $\SetZ^r_v$ to the set of rows of a
  $\voa(M,v)$ $\mb{T}$.
\end{lemma}
\begin{proof}
 For any $A\subseteq N$, let $\mb{U}(A)$ be the submatrix of $\mb{U}$ whose
 columns are those of $\mb{U}$ indexed by $A$.
  Let $B\subseteq N$ be a basis of $M$.  Then $\det(\mb{U}(B))=1$ or $-1$, and by
  Lemma \ref{lr}, each row of $\mb{T}(B)$ occurs in $\mb{T}(B)$
  exactly once.

  Let $I$ be any independent set of $M$. There exists a
  basis $B$ such that $I\subseteq B$. Thus $\mb{T}(I)$ can be
  considered a subarray of $\mb{T}(B)$, thus each row of  $\mb{T}(I)$
  occurs in $\mb{T}(I)$ $v^{\br(N)-|I|}$ times.
 
Now for any $A\subseteq N$, let $I\subseteq A$ be a maximal
independent set contained in $A$. Thus, $\br(A)=|I|$. According to \cite[Lemma
2.2.20]{O11}, by pivoting a non zero entry in $\mb{U}(A)$, it can be
written as 
\begin{equation*}
\begin{bmatrix}
  \mb{I}_r & \mb{D} \\
  \mb{0}_{(\br(N)-r)\times r}  & \mb{0}_{(\br(N)-r)\times( |A|-r)}
\end{bmatrix},
\end{equation*}
where $r=|\br(A)|$, $\mb{I}_r$ is the $r\times r$ identity matrix and $\mb{D}$ is a
$r\times (|A|-r)$ totally unimodular matrix,
$\mb{0}_{(\br(N)-r)\times r}$  and $\mb{0}_{(\br(N)-r)\times( |A|-r)}$ are zero matrices. Hence, columns of $\mb{U}(A\setminus
I)$ linearly depend on columns of $\mb{U}(I)$, and therefore each
row $\mb{b}$ of $\mb{T}(A)$ occurs the same times as the row
$\mb{b}(I)$ occurs in $\mb{T}(I)$, i.e.,
$v^{\br(N)-|I|}=v^{\br(N)-\br(A)}$ times. The lemma follows.
\end{proof}

By Lemma \ref{voazv}, we immediately obtain the following proposition.

\begin{proposition}
    \label{regular}
  For any regular matroid $M$, $\chi_M=\{v\in\mathbb{Z} : v\ge 2\}$.
\end{proposition}



Now we have determined the p-characteristic set of a regular
matroid, i.e., the set of all integers exceeding one. We will
prove in the following that the converse of the proposition is also
true, that is, the matroid whose p-characteristic set is the set of
all integers exceeding one must be regular. We state it in
Proposition \ref{regular1}. Note that the excluded minors of the
classes of regular matroids are $U_{2,4}$, the Fano matroid $F_7$ and
the dual of Fano matroid $F_7^*$ \cite[Theorem 10.1.1]{O11}, which
means for any matroid that is not regular must contain one of these
\rv{three} matroids as a minor. Then by Theorem  \ref{minor}, to prove Proposition \ref{regular1}, it suffices to prove that none of p-characteristic set of 
 these three matroids is equal to the set of all integers greater or
 equal to 2. For $U_{2,4}$, it is well-known and stated in
 \cite[Proposition 2]{CCB21} that $\chi_{U_{2,4}}=\{v\in\SetZ: v\ge 3,
 v\neq 6\}$. For $F_7$, it is proved in \cite[Proposition 4.1]{M99}
 that $v\in \chi_{F_7}$ if and only if $v$ is a power of $2$. Now we
 will prove in Lemma \ref{df7} the p-characteristic set of $F^*_7$
\rv{contains not all integers exceeding one}. For
 $F^*_7=(N, \br)$, it is a matroid with $N=\{1,2,\cdots, 7\}$ and
 \begin{equation*}
   \br(A)=
   \begin{cases}
     \ |A|, \quad |A|\le 3,\\
     \ 3,\quad A\in\{\{1,2,3,5\},\{1,2,4,6\},\{1,3,4,7\},\\ \quad \quad\{1,5,6,7\},\{2,3,6,7\},\{2,5,4,7\},\{3,4,5,6\}\}\\
     \ 4,\quad \text{o.w.}
   \end{cases}
 \end{equation*}
for any $A\subseteq N$.

\begin{lemma}
\label{df7}
  For $F^*_7$, the dual matroid of the Fano matroid $F_7$, $3\not\in\chi_{F^*_7}$.
\end{lemma}

\begin{proof}
  Let $F^*_7$ be defined above, where $B=\{1,2,3,4\}$ is a basis. Now
  we try to construct a $\voa(F^*_7, 3)$ $\mb{T}$ in the
  following. Let each $\mb{a}\in \SetZ^4_3$ be a row in
  $\mb{T}(B)$. We construct column $5$ (resp. $6$ and $7$)
of $\mb{T}$ such that $\ded(\mb{T}(\{1,2,3,5\}))$
(resp. $\ded(\mb{T}(\{1,2,4,6\}))$ and  
$\ded(\mb{T}(\{1,3,4,7\}))$ ) forms an $\oa(3,4,3)$.
Note that there exist $24$ distinct $\oa(3,4,3)$s (up to row
  permutation).\footnote{Each $\oa(3,4,v)$ (up to row
  permutation) corresponds to a latin cube of order $v$ which has 24
  types when $v=3$, see \cite{OeisA098679}.}  For the all $24^3$ possible chosen
of $\mb{T}(\{5,6,7\})$, we check the all $24^3$
candidates of $\mb{T}$ \rv{with a MATLAB program} by computer.
It turns out that at least
 one of $\ded(\mb{T}(\{1,5,6,7\}))$, $\ded(\mb{T}(\{2,3,6,7\}))$, $\ded(\mb{T}(\{2,5,4,7\}))$ and $\ded(\mb{T}(\{3,4,5,6\}))$
does not form a $\oa(3,4,3)$. Thus, $\voa(F^*_7, 3)$ is not
constructible and therefore $3\not\in\chi_{F^*_7}$.
\end{proof}

By Lemma \ref{df7} and the discussion above, we immediately have
Proposition \ref{regular1}.

\begin{proposition}
    \label{regular1}
  For a matroid $M$, if $\chi_M=\{v\in\mathbb{Z} : v\ge 2\}$, then $M$
  is a regular matroid.
\end{proposition}

Propositions \ref{regular} and \ref{regular1} yield the following
theorem.

\begin{theorem}
For a matroid $M$, $\chi_M=\{v\in \mathbb{Z} : v\ge 2\}$ if and only
if $M$ is regular.
\end{theorem}

By the discussion in \cite[Chapter 6]{O11}, we know that regular matroids are a large class
of matroids which include graphic matroids, cographic matroids and
matroid $R_{10}$ (See \cite[Section 6.6]{O11} for its
definition). Thus, the task we completed in this section also
let us know the p-characteristic set of these subclass of matroids.


\subsection{Matroids with the same p-characteristic set as
  $U_{2,4}$}
\label{chu24}

By Corollary \ref{3con}, to determine the
p-characteristic set of a matroid, we only need to consider those
smallest building blocks, i.e., $3$-connected matroids. In this
subsection, we first consider a class of 3-connected matroids, that is,
whirl matroids $\mc{W}^r$.

\begin{figure}[h]
       \centering
       \includegraphics[width=6cm]{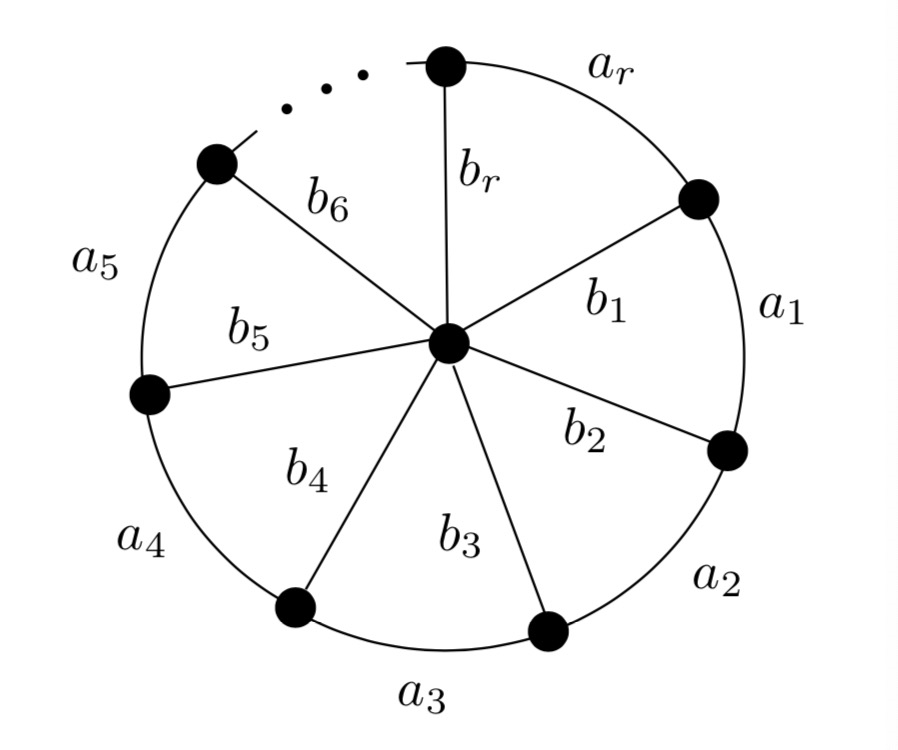}
       \caption{The wheel graph $\mc{W}_r$}
       \label{wheel}
\end{figure}

The cycle matroid of the graph in Figure \ref{wheel} is called the \emph{wheel
matroid} $M(\mc{W}_r)$. It is a matroid with ground set $N=A\cup B$, where
$A=\{a_i:i=1,2,\cdots,r\}$ called rim of $M(\mc{W}_r)$ and
$B=\{b_i:i=1,2,\cdots,r\}$ called the set of spokes, and
family of circuits $\mc{C}$ being the family of the cycles of
$\mc{W}_r$. 

The whirl matroid $\mc{W}^r$ is a matroid by relaxing the 
circuit-hyperplane $A$, i.e., the rim of the wheel matroid
$M(\mc{W}_r)$ (see \cite[Example 1.5.13 and Proposition 1.5.14]{O11}
for the relaxing of a circuit-hyperplane of a matroid). Thus, $\mc{W}^r=(N,\mc{C})$
with $N=A\cup B$ and family of circuits

\begin{equation*}
  \mc{C}=(\mc{C}(M(\mc{W}_r))-A)\cup \{A\cup b_i:1\le i\le r\}.
\end{equation*}
It can be seen that when $r=2$, $\mc{W}^r=U_{2,4}$. For more about
wheel and whirl matroids, readers are referred to \cite[Section 8.4]{O11}.

\begin{proposition}
  \label{pw}
  For matroid $\mc{W}^r, r\ge2 $, $\chi_{\mc{W}^r}=\chi_{U_{2,4}}=\{v\in
\SetZ: v\ge 3, v\neq 6\}$.
\end{proposition}
\begin{proof}
  Note that $\mc{W}^{r-1}=\mc{W}^{r}/a_r\setminus b_r$, so $\mc{W}^{r-1}$ is a
  minor of $\mc{W}^{r}$ for any $r$. As a minor's minor is a minor,
  $\mc{W}^2=U_{2,4}$ is a minor of $\mc{W}^{r}$ for any $r$.
  By Theorem \ref{minor},
  $\chi_{\mc{W}^r}\subseteq \chi_{U_{2,4}}$. In the following, we
  prove another direction of inclusion by induction.

As $\mc{W}^2=U_{2,4}$, we only
need to prove that $\chi_{\mc{W}^{k+1}}\subseteq \chi_{\mc{W}^k}$ for
any $2\le k \le r-1$. Now assume $\mb{a}_i$ and $\mb{b}_i$, $1\le i\le k$ are
the columns of a $\voa(\mc{W}^k,v) $ indexed by $a_i\in A$ and
spokes $b_i\in B$, $1\le i\le k$, respectively. Now we construct a
$v^{k+1}\times (2k+2)$ array $\mb{T}$ with columns $\mb{a}'_i$ and
$\mb{b}'_i$, $1\le i\le k+1$ in the following way. 
Let
\begin{equation*}
  \mb{a}'_i(tv^k+j)=
  \begin{cases}
   \ \mb{a}_i(j)\quad &1\le i\le k-1\\
   \ \mb{a}_i(j)\otimes_1 t \quad &i=k\\
   \ t\quad &i=k+1.
  \end{cases}
\end{equation*}
and
\begin{equation*}
  \mb{b}'_i(tv^k+j)=
  \begin{cases}
   \ \mb{b}_i(j)\quad &1\le i\le k\\
    \ \mb{b}_k(j)\otimes_2 \mb{a}'_k(j) \quad &i=k+1
  \end{cases}
\end{equation*}
where $j=0,1,\cdots,v^k -1$ and $t=0,1,\cdots, v-1$, and $\otimes_i,i=1,2$
are defined in the following.

Let $\mb{T}_1$ be an arbitrary $\oa(2,3,v)$. For any $c_1,c_2\in
\SetZ_v$, $c_1\otimes_1c_2\in \SetZ_v$ is defined such that $(c_1,c_2,
c_1\otimes_1c_2 )$ forms a row of $\mb{T}_1$. 
Let $\mb{T}_2$ be an
array formed by columns $\mb{a}'_k$, $\mb{b}'_k$, $\mb{b}'_1$ and
$\mb{a}'_{k+1}$. Let $\mb{T}'_2=\ded(\mb{T}_2)$. 
 It can be checked that
$\mb{T}'_2$  is an $\oa(3,4,v)$ or a $\voa(U_{3,4},v)$. 
Let $\mb{T}_3=\mb{T}'_2|_{a_{k+1}:0}$. Thus by Proposition
\ref{pcont}, $\mb{T}_3$ is a $\voa(U_{3,4}/a_{k+1},v)$ or a
$\voa(U_{2,3},v)$ or a $\oa(2,3,v)$.  For any $c_1,c_2\in
\SetZ_v$, $c_1\otimes_2c_2\in \SetZ_v$ is defined such that $(c_1,c_2,
c_1\otimes_2c_2 )$ forms a row of $\mb{T}_3$.

It can be checked that $\mb{T}$ is an $\voa(\mc{W}^{k+1}, v)$, which
implies the proposition. 
\end{proof}


By Corollary \ref{3con}, Proposition 
\ref{regular} and Proposition
\ref{pw}, we immediately obtain the following theorem.

\begin{theorem}
  \label{tu24}
  For any matroid $M$, let $M_i$ be its connected components,
  and $M_{i,j}$ be the 3-connected components of $M_i$. Then
  $\chi_{M}=\chi_{U_{2,4}}$ if each of these $M_{i,j}$ is either a regular
  matroid or a $\mc{W}^r$ with $r\ge 2$, and at least one of them is a  $\mc{W}^r$.
  
\end{theorem}

  \rv{Note that Theorem \ref{tu24} provides a sufficient condition for
  us to determine whether a matroid has the same p-characteristic set
  as $U_{2,4}$. We will discuss in Subsection \ref{extrac} that such a matroid
  should has no minors $U_{2,5}$, $U_{3,5}$, $F_7$ and $F^*_7$ and not
  regular, which is a necessary condition.
  } 

\section{Discussion}
\label{disc}

\subsection{Entropy functions on the extreme rays of polymatroidal
  region}
\label{extray}

One motivation of the study of matroidal entropy function is to
characterize entropy functions on the extreme rays of polymatroidal
regions. 
\rv{As $\Gamma_N$ is a polyhedral cone defined by a set of
  inequalities which can be reduced to the minimal characterization
  called elemental inequalities (cf. Chapter 14 in \cite{Y08}), an extreme ray of
  $\Gamma_N$ is the intersection of the hyperplanes determined by
  setting some subset of elemental inequalities to equalities. Note
  that the coefficients of elemental inequalities are $0, 1$ or $-1$,
  polymatroids on an extreme ray can be considered as the solutions to
  a system of 
  linear equations with such coefficients, and thus there must
  be integer polymatroids on the ray, that is, $\h\in\Gamma_N$ with $\h(A)$ integer for all $A\subseteq N$. Among
  the integer polymatroids an extreme ray contains,
  there must be a unique minimal $\h$, that is, for any integer $t\ge 1$,
  $\h/t$ is integer if and only if $t=1$.  
} 

\rv{We first consider those extreme rays of $\Gamma_n$ whose
 minimal integer polymatroid is a matroid.}
It has been proved in \cite{N78}
that the \rv{rank function} of a matroid is on an extreme ray of the
polymatroidal region $\Gamma_N$ if and only if it is connected after loops
deleted. Thus, for a ($2$-)connected matroid or a matroid $M$ obtained by adding
loops to a connected matroid with rank exceeding one, the \rv{extreme ray}
of $\Gamma_N$ containing $\br_M$ is fully characterized by
its p-characteristic set $\chi_M$. That is, for any $\h= a\cdot \br_M$ on
the extreme ray of $\Gamma_N$ containing $\br_M$, $\h$ is entropic if
and only if $a=\log v$ and $v\in\chi_M$.
For an extreme ray of
$\Gamma_N$ containing matroid $M$ with rank $1$, by the discussion in the Remark
after Theorem \ref{thm}, the whole ray is entropic, i.e., all $\h$ on
the ray belong to $\Gamma^*_N$.

For an extreme ray of $\Gamma_N$, if the minimal
integer polymatroid it contains is not (the rank function of) a
matroid, we can characterize it by a \emph{mixed level variable strength
  orthogonal array}(MVOA) which is a generalization of $VOA$. In the
following, we distinguish an integer polymatroid $P=(N,\br)$ from
its rank function $\br$.

\begin{definition}[MVOA]
  Given a loopless integer polymatroid $P=(N,\mb{r})$ with $\br(N)\ge 2$, a $v^{\br(N)}\times n$
  array $\mb{T}$ with columns indexed by $N$,
  whose entries of a column indexed by $i\in N$
 are from $\{0,1,\ldots, v^{\br(i)}-1\}$, is called a \emph{mixed level variable strength}
  orthogonal array(MVOA) induced by $P$
  with \emph{base level} $v$ if for each
 $A\subseteq N$, $v^{\mb{r}(N)}\times |A|$ subarray $\mb{T}(A)$ of $\mb{T}$ consisting of
 columns indexed by $A$ satisfies the following condition:
each row of  $\mb{T}(A)$ occurs in $\mb{T}(A)$ exactly $v^{\br(N)-\br(A)}$ times.
 We also call such $\mb{T}$ a $\mvoa(P,v)$.
\end{definition}

\rv{Note that when $P$ is a matroid, i.e., $\br(i)=1$ for all $i\in
  N$, an $\mvoa(P,v)$ becomes a VOA.} 
Similar to the p-characteristic set of a matroid, we can also define
the p-characteristic set of an integer  polymatroid $P=(N,\mb{r}_P)$  by
\begin{equation*}
  \chi_P=\{v\in \SetZ: v\ge 2, \log v\cdot \br_P\in \Gamma^*_N\}.
\end{equation*}

\begin{definition}[free expansion]
\rv{  For a polymatroid $P=(N,\br)$, let $N'$ be a set with
  cardinality $\sum_{i\in N}\br(i)$. Let $\phi: N\to 2^{N'}$
  be a mapping such that each $\phi(i)$ has cardinality $\br(i)$ and
  they are disjoint. Then the \emph{free expansion} of $P$ by $\phi$ is
  defined by }

  \begin{equation*}
    K \mapsto \min_{I\subseteq K} \br(I)+|K\setminus \phi(I)|.
  \end{equation*}
\end{definition}

\rv{See} \cite[Theorem 1.3.6]{N78} \rv{that the free expansion of an
  integer polymatroid $P$ is \rvv{well-defined}, and unique up to isomorphism.} 
It is not difficult
to check that $\chi_M \subseteq\chi_P$, where matroid $M$ is the free
expansion of $P$. However, the following example shows
that there exists integer polymatroids $P$ such that $\chi_M\subsetneq \chi_P
$.
\begin{example}
  Let $\hat{U}_{2,5}$ be the integer polymatroid on $N=\{1,2,3,4\}$
  whose rank function is
  defined by
  \begin{equation*}
    \h(A)=
     \begin{cases}
       \ \min\{2,|A|\}\quad &A\neq \{1\}\\
      \  2\quad &A=\{1\}.
  \end{cases}
  \end{equation*}
 It can be seen that its free expansion is the uniform matroid
 $U_{2,5}$ \rv{on $N'=\{1,1',2,3,4\}$, where the mapping $\phi$ defined by
 $\phi(1)=\{1,1'\}$ and $\phi(i)=\{i\}, i=2,3,4$}. It is well known that $2\notin \chi_{U_{2,5}}$. However, the
 following $\mvoa(\hat{U}_{2,5},2)$ shows that $2\in \chi_{\hat{U}_{2,5}}$.
   \begin{equation*}
     \begin{matrix}
     0 & 0 & 0 & 0 \\
     1 & 0 & 1 & 1 \\ 
    2 & 1 & 0 & 1 \\
     3 & 1 & 1 & 0 
   \end{matrix}
 \end{equation*}
 Thus, $\chi_{U_{2,5}}\subsetneq \chi_{\hat{U}_{2,5}}$
\end{example}
Nevertheless, the p-characteristic set of its free expansion provide
an inner bound of the p-characteristic of an integer polymatroid.

\begin{question}
 Let $P$ a polymatroid, and $M$ its free expansion. In which condition, $\chi_P=\chi_M$?
\end{question}

By matroidal entropy functions, we provide an approach to characterize extreme rays, or
$1$-dimensional face of $\Gamma_N$. In \cite{LC23, LCC24} 
we also try to characterize $2$-dimensional face of $\Gamma_N$.

\subsection{Connectivity}

For a matroid $M$ which is the direct sum of connected matroids $M_i$,
where $M_i=(N_i,\br_{M_i})$, the ray containing its rank function
$\br_M$ is on the face \footnote{\rvv{To show $M$ is on a face of
    $\Gamma_N$, we write $\br_M=\sum^t_{i=1}\br_{M'_i}$, where each $M'_i$
    is a matroid on $N$ obtaind by adding loops to $M_i$ on
    $N\setminus N_i$. Then it must be
    on the face $\{\h\in\Gamma_N: \h(N)=\sum^t_{i=1}\h(N_i)\}$ of
    $\Gamma_N$, where $\h(N) \le  \sum^t_{i=1}\h(N_i)$ is a Shannon-type
    inequality.} } of $\Gamma_N$ which is the conic hull of the
rank functions of $M'_i$, where $M'_i$ is a matroid on $N$ obtained by adding loops in
$N\setminus N_i$ to $M_i$. By Theorem \ref{tdsum}, for $\h=a\cdot
\br_M$ is entropic if and only if $a=\log v$ and $v\in\chi_M=\cap_i\chi_{M_i}$. 

In Theorem \ref{t2sum}, we proved a similar properties of 
the matroid operation
$2$-sum of a matroid and its $3$-connected
components. Now it is natural to ask the following question.

\begin{question}
  For a $2$-connected matroid and its $3$-connected components, how
  are they geometrically related on $\Gamma_N$?
\end{question}

\subsection{Excluded minor and classification of matroids according to its p-characteristic set}
\label{extrac}

The definition of the minors of a matroid yields a partial
order on the set of all matroids. Let $\mc{M}$ be the set of all finite
matroids. For the partial order $(\mc{M, \preceq})$, for any
$M_1,M_2\in \mc{M}$, $M_1\preceq M_2$ if $M_1$ is a minor of
$M_2$. For a class $\mc{M}'\subseteq \mc{M}$, a matroid $M$ is called an
\emph{excluded minor} of $\mc{M}'$ if $M\not\in \mc{M}'$ and each proper
minor of $M$ is in $\mc{M}'$. Thus the set of excluded minors of
$\mc{M}'$ coincides the family of minimal elements of the partial order set
$(\mc{M}\setminus \mc{M}',\preceq)$.

On the other hand, such $\mc{M}'$
must be \emph{minor-closed}, that is, if $M\in \mc{M}'$, each minor
$M'$ of $M$ must be in $\mc{M}'$. It can be seen from Subsection
\ref{srm} that the family of regular matroids, which is also the the family of
matroids with the p-characteristic set $\{v\in \SetZ: v\ge
2\}$, is minor-closed. Thus, it can be determined by its excluded minors $U_{2,4}$,
$F_7$ and $F^*_7$. However, in general, a class $\mc{C}$ of all matroids with
its p-characteristic set a specific $\chi_{\mc{C}}$ may not be
minor-closed. Nonetheless, it is implied by Theorem \ref{minor} that the
mapping $f:\mc{M}\to \mc{P}$, where $\mc{P}$ is a partially ordered
set on the family of all subsets of $\{v\in \SetZ: v\ge
2\}$ and ordered by inclusion, is monotonic,  so
the class $\mc{C}'$ of all matroids with p-characteristic
$\chi'\supseteq \chi_{\mc{C}}$,
is minor-closed.

\begin{lemma}
\label{exc}
For any $i\ge 2$ and $i\neq 3$, the set $\chi_i\triangleq\{v\in \SetZ: v\ge2, v\neq i\}$ is not
the p-characteristic set of any matroid.
\end{lemma}
\begin{proof}
  Assume the lemma is not true, and $M$ is a minimal matroid with
  $\chi_M=\chi_i$ in the sense that for any proper minor $M'$ of $M$,
  $\chi_{M'}\neq \chi_M$. By Theorem \ref{minor}, $\chi_{M'}\supseteq
  \chi_i$, and so $\chi_{M'}=\{v\in \SetZ: v\ge2\}$. Then by Proposition
  \ref{regular1}, $M'$ is a regular matroid. Thus, $M$ is an excluded
  minor of the family of all regular matroid. It contradicts the fact
  that such excluded minor can only be $U_{2,4}$, $F_7$ and $F^*_7$.
\end{proof}

Let $\mc{M}_{\rm reg}$ be the family of all regular matroids and
$\mc{M}_{U_{2,4}}$ be the family of all matroids with the same
p-characteristic set as $U_{2,4}$, i.e. $\{v\in \SetZ: v\ge
3, v\neq 6\}$. By Lemma \ref{exc}, there does not exist a matroid whose
p-characteristic set is $\chi_2$ or $\chi_6$,
we immediately have the following proposition. 

\begin{proposition}
  \label{cfu24}
 The family of matroids $\mc{M}'_{U_{2,4}}=\mc{M}_{\rm reg}\cup \mc{M}_{U_{2,4}}$ is
 minor-closed. 
\end{proposition}

With Proposition \ref{cfu24}, one naturally asks the following question.

\begin{question}
  What's the set of excluded minors of $\mc{M}'_{U_{2,4}}$?
\end{question}

It is not difficult to check that $U_{2,5}$, $U_{3,5}$, $F_7$ and
$F^*_7$ are excluded minors of $\mc{M}'_{U_{2,4}}$. Whether they form
the full list needs to be further proved.

\rv{Thus, if a matroid has the
same p-characteristic set as $U_{2,4}$, then it has no minors
$U_{2,5}$, $U_{3,5}$, $F_7$ or 
$F^*_7$, and it is not regular. }

\subsection{Duality}
\label{dul}
Duality between matroids may also be useful to classify the matroids
according to its p-characteristic set. It has been proved in \cite[Theorem
2.2.8]{O11} that if a matroid $M$ is representable over a field
$\mathbb{F}$, its dual $M^*$ is also representable over $\mathbb{F}$. The
result holds since if $M$ is the vector matroid of an $r\times n$ matrix $\mb{A}$
with rank $r$,
then $M^*$ is the vector matroid of the $(n-r)\times n$ $\mb{A}^\perp$
with rank $n-r$, the orthogonal
matrix of $\mb{A}$.

Now consider $\mathbb{F}=\mr{GF}(q)$ for some prime power $q$. Note that the mapping $\mb{x}\mapsto \mb{x}\mb{A},
\mb{x}\in \SetZ_q^r$ maps $\SetZ_q^r$ to the family of rows of a
$\voa(M, q)$ and $\mb{y}\mapsto \mb{y}\mb{A}^{\perp},
\mb{y}\in \SetZ_q^{n-r}$ maps $\SetZ_q^{n-r}$ to the family of rows of a
$\voa(M^*, q)$. The family of rows of a $\voa(M, q)$ forms the code
book $\mc{C}$ of a linear code, while the family of rows of the $\voa(M^*, q)$
forms the book $\mc{C}^*$ of the dual code of $\mc{C}$. 
Thus, $q\in\chi_M$ if and only if $q\in \chi_{M^*}$. 

Now for a general integer $v\ge 2$, if a $\voa(M,v)$ $\mb{T}$ exists,
can we construct a $\voa(M^*, v)$ according to $\mb{T}$? If so, we can
prove the following conjecture. 

\begin{conjecture}
\label{cex}
  For a matroid $M$ and its dual $M^*$, $\chi_{M^*}=\chi_M$.
\end{conjecture}

Note that matroid operations deletion and contraction, series
and parallel connections form dual operations, and direct sum and
$2$-sum are self-dual. Our discussion in Section \ref{con} provides the
corresponding VOA constructions of these matroid operations, which
reals some clues of Conjecture \ref{cex}.

\begin{question}
  Prove or disprove Conjecture \ref{cex}. 
\end{question}

As we discussed in \cite[Propositions 3 and 4]{CCB21}, for $U_{2,5}$,
the only undecided element whether in $\chi_{U_{2,5}}$ or not is $10$.
However for $U_{3,5}$, the dual of $U_{2,5}$, for $v\equiv 2\pmod 4$,
only a small amount of them can be decided that they are in
$\chi_{U_{3,5}}$. (See \cite{JY10, YWJL11}  for more details of the
best result so far of this problem.) Hence, if we can prove Conjecture
\ref{cex}, more about $\chi_{U_{3,5}}$ will be known from
$\chi_{U_{2,5}}$.

\subsection{Representations}

In \cite{M99}, Mat\'u\v{s} defined partition representation of a
matroid. As we discussed in Section \ref{sec:1}, each row of a
$\voa(M,v)$ can be considered as the label of each block of the partition
representation of $M$. Partition representation is a generalization of
the linear representation of a Matroid. 
In the previous subsection, for a representable matroid $M$
over $\mr{GF}(q)$, we can construct a $\voa(M,q)$ by mapping $\mb{x}\mapsto \mb{x}\mb{A},
\mb{x}\in \SetZ_q^r$. In the proof of Lemma \ref{voazv}, we do the
mapping analogously but by a matrix over ring $\SetZ_v$. Then it is
natural to extend this problem the matrix over a general ring.

\begin{question}
  In which condition, a $\voa(M,v)$ can be generated by a matrix over a
  ring, that is, $M$ is representable over a ring? 
\end{question}


\subsection{Applications to network coding}
\label{anc}
\rv{
As we discussed in Section I, matroidal entropy functions and the
corresponding VOAs can be
applied to construct codes in information theory problems such as
network coding, secret sharing, index coding and locally repairable
codes. In this subsection, we gave a brief introduction of the
applications to network coding.

Let $G=(\mc{V},\mc{E})$ be an acyclic directed multigraph, where $\mc{V}$
is the set of vertices and $\mc{E}$ the set of edges. Let
$\mc{S}\subset\mc{V},\mc{T}\subset\mc{V}$ be the set of sources and
set of sinks, respectively, and assume
$\mc{S}\cap\mc{T}=\emptyset$. For each $e\in\mc{E}$, let $R_e$ be the
capacity of $e$. For each $t\in\mc{T}$, let $\alpha(t)\subseteq\mc{S}$
be the set of sources it needs to receive.  
For each source $s\in\mc{S}$, we assume it has an
imaginary incoming edge. Let $N=\mc{E}\cup\mc{S}$. For a vertex $v\in
\mc{V}$, let $\mr{In}(v)$ and $\mr{Out}(v)$ the incoming and outcoming
edges of $v$, respectively. 
Then the support of  
random vector $X_N=(X_i:i\in N)$ forms a network code if 
}
\begin{itemize}
\item \rv{for each $v\in\mc{V}$, and $e\in \mr{Out}(v)$, there exists
    an encoding function $k_e$ such that $X_e=k_e(X_i:i\in \mr{In}(v))$;}
\item \rv{for each $e\in \mc{E}$, $H(X_e)\le R_e$;} 
\item \rv{for each $t\in \mc{T}$, and each $s\in\alpha(t)$, there
    exists a decoding function $g_{t,s}$ such that
    $X_s=g_{t,s}(X_i:i\in \mr{In}(t))$.}
\end{itemize}
\rv{Now, if we set $R_e=1$ for all $e\in\mc{E}$ and let $H(X_i)=1$ for
  all $i\in N$. Then entropy function of $X_N$ will be a matroidal
  entropy function induced by a matroid $M$ on $N$, and $M$ satisfies
}
\begin{itemize}
\item \rv{ for each $v\in\mc{V}$, and each $e\in \mr{Out}(v)$, 
    $D_e=\{e\}\cup \mr{In}(v)$ is dependent and there exists a circuit
    $C_e\subseteq D_e$ containing $e$; }
\item \rv{for each $t\in \mc{T}$, and each $s\in\alpha(t)$,
    $D_s=\{s\}\cup \mr{In}(t)$ is dependent and there exists a circuit
  $C_s\subseteq D_s$ containing $s$.}
\end{itemize}
\rv{Therefore, we could use the corresponding $\voa(M,v)$ for some
$v\in\chi_M$ to construct a network coding scheme. If such
a $\voa(M,v)$ exists, we say the network is \emph{solvable}. 
For some prime power $q$, if there exists a coding scheme $\voa(M,q)$
which is a linear code over $\mathbb{F}_q$, we say that this network
is \emph{linearly solvable} over $\mathbb{F}_q$. }

\rv{The following example is from \cite{DFZ04}.}

          \begin{figure}[h]

	\begin{center}
		\includegraphics[width=2.3in]{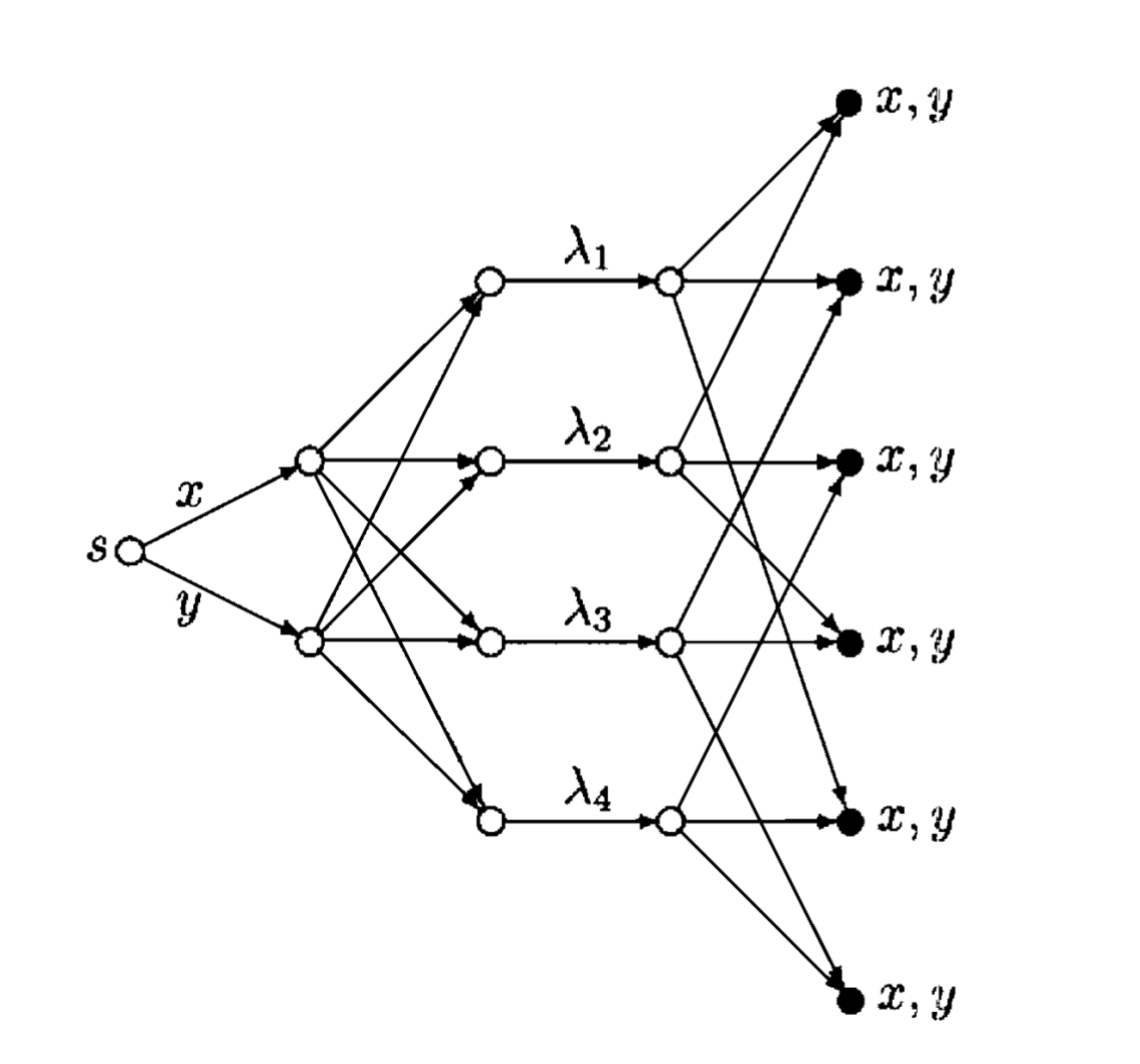}
              \end{center}
              \caption{The $(4, 2)$-combination network}
              \label{com}
      \end{figure}

\begin{example}
\rv{Figure \ref{com} shows a network with $\mc{S}=\{x,y\}$ and 
the set $\mc{T}$ of solid vertices. For each $t\in\mc{T}$,
$\alpha(t)=\{x,y\}$. This network is called $(4, 2)$-combination network.
One coding scheme is
$\lambda_1=x$, $\lambda_2=y$, $\lambda_3=L_1(x,y)$,
                $\lambda_4=L_2(x,y)$, where $L_1,L_2$ are the symbols
                of the mutually orthogonal latin squares. Thus,
                $\{\lambda_1$, $\lambda_2$,$\lambda_3$,$\lambda_4\}$ forms
                $\voa(U_{2,4},v)$. Other edges of the network can be
                considered as parallel edges of
                $\lambda_i,i=1,2,3,4$. Thus matroid of the network can
                be considered as the one whose simplification is $U_{2,4}$.}
\end{example}

\rv{
Besides \cite{DFZ04}, there exist a number of literatures on this
topic. For example, 
in \cite{SYLL15}, authors constructed the first known single-source multicast network called
the Swirl network that is linearly solvable over a small field but not
over a larger field, which is induced by Swirl matroid. 
In \cite{LS01}
the duality property of a representable matroid was adopted to
construct convolutional network codes over a network with cycles. 
In \cite{TS19}, an equivalence among the vector linear solvability of a
network coding problem, the vector linear solvability of an index
coding problem and the representability of a discrete polymatroid was
established.

}


\begin{thebibliography}{1}
\bibitem{CCB22}
Q. ~Chen, M. ~Cheng and B. ~Bai, ``Matroidal entropy functions:
characterizations, constructions and representations,'' IEEE
Int. Symp. Info. Theory, Espoo, Finland,  June
  2022.
 
\bibitem{F78}
S. Fujishige, ``Polymatroidal dependence structure of a set of
random variables,'' \emph{Info. Contr.}, vol. 39: pp. 55-72,1978.

\bibitem{ZY98}
Z. Zhang and R. W. Yeung, ``On characterization of entropy function
via information inequalities,'' \emph{IEEE Trans. Info. Theory}, vol.
  44, pp. 1440-1452, Nov. 1998.


\bibitem{YYZ01}
X. Yan, R. W. Yeung and Z. Zhang, ``A class of non-Shannon-type
information inequalities and their applications,'' IEEE Int. Symp. Info. Theory, Washington DC, June
  2001.

\bibitem{MMRV02}
K.~Makarychev, Y.~Makarychev, A.~Romashchenko, et al, ``A new class of
non-Shannon-type inequalities for entropies,'' \emph{Communications in
Information and Systems}, vol. 2, no. 2, pp. 147-166, 2002


\bibitem{Z03}
Z. ~Zhang ``On a new non-Shannon-type information
inequality,'' \emph{Communications in Information and Systems,}
vol. 3, no. 1, pp. 47-60, June 2003.


\bibitem{DFZ06}
R. ~Doughterty, C. ~Freiling and K. ~Zeger, ``Six new non-Shannon
information inequalities,'' IEEE Int. Symp. Info. Theory, Seattle WA June
  2006.


\bibitem{M07a}
F.~Mat\'u\v{s}, ``Infinitely many information inequalities,'' 
IEEE Int. Symp. Info. Theory, Nice, France, June
  2007.
  

\bibitem{CCB21}
Q.~Chen, M.~Cheng and B.~Bai, ``Matroidal entropy functions: a quartet
of theories of information, matroid, design and coding,''
\emph{Entropy,} vol. 23:3, 1-11, 2021.

\bibitem{M99}
F.~Mat\'u\v{s}, ``Matroid representations by partitions,''
\emph{Discrete Math.} vol. 203 pp. 169–194, 1999


\bibitem{O11}
J. G. Oxley, \emph{Matroid theory,} 2nd ed. Oxford Univ. Press, 2011.



\bibitem{SA98}
J. Simonis and A. Ashikhmin, ``Almost affine codes,''\emph{Designs, Codes
  Cryptogr., } vol. 14, pp. 179–197, 1998.


\bibitem{BD91}
E. F. Brickell.; D. M. Davenport, ``On the classification of ideal
secret sharing schemes,''  \emph{J. Cryptol.}  vol. 4, 123-134, 1991.

\bibitem{DFZ07}
 R. Dougherty, C. Freiling and K Zeger,``Networks, matroids, and
  non-Shannon information inequalities,'' \emph{IEEE
    Trans. Inf. Theory}  vol. 53, no. 6, pp. 1949-1969, 2007.
    
\bibitem{DFZ04}
  R. Dougherty, C. Freiling and K Zeger, ``Linearity and solvability in multicast networks,''  \emph{IEEE
    Trans. Inf. Theory}  vol. 50, no. 10, pp. 2243-2256, 2004.


\bibitem{SYLL15}
Q. T. Sun, X. Yin, Z. Li and K. Long, 
  ``Multicast network coding and field sizes,''  \emph{IEEE
    Trans. Inf. Theory}  vol. 61, no. 11, pp. 6182-6191, 2015.
        
         
\bibitem{LS01}
S.-Y. R. Li and Q. T. Sun,
  ``Network coding theory via commutative algebra'', \emph{IEEE
    Trans. Inf. Theory}  vol. 57, no. 1, pp. 403-415, 2011.


\bibitem{TS19}
A. Thomas and S. Rajan,
  ``Linear network coding, linear index coding and
  representable discrete polymatroids,''
 \emph{IEEE
    Trans. Com.}  vol. 67, no. 7, pp. 4593-4604, 2019.
 
\bibitem{MS16}
V. T. Muralidharan and B. S. Rajan,
  ``A discrete polymatroidal framework for differential
  error-correcting index codes,''  \emph{IEEE
    Trans. Inf. Theory}  vol. 62, no. 7, pp. 4096-4119, 2016.

\bibitem{RSG10}
S. El Rouayheb, A. Sprintson and C. Georghiades, ``On the index coding problem and its relation to
network coding and matroid theory'', \emph{IEEE
  Trans. Inf. Theory}  vol. 56, no.7  pp. 3187-3195, 2010.

\bibitem{WFEH16}
 T. Westerb\"{a}ck, R. Freij-Hollanti, T. Ernvall and C. Hollanti,
 ``On the combinatorics of locally repairable
codes via matroid theory'', \emph{IEEE
    Trans. Inf. Theory}  vol. 62, no.10  pp. 5296-5315, 2016.

  \bibitem{HSS99}
A. S. Hedayat, N. J. A. Sloane,. J. Stufken, \emph{Orthogonal Arrays: Theory and
 Applications};  Springer: New York, NY, USA, 1999.

\bibitem{B93}
  W. C. Brown, \emph{Matrices over commutative rings,} Marcel
  Dekker, INC., 1993.


\bibitem{OeisA098679}
N. J. A. Sloane,  ``Total number of Latin cubes of order n'', 
The On-Line Encyclopedia of Integer Sequences,
https://oeis.org/A098679, Nov. 2004. 





\bibitem{Y08}
R. W. Yeung, \emph{Information Theory and Network Coding}, Springer, 2008. 

\bibitem{N78}
H. Q. Nguyen, ``Semimodular functions and combinatorial geometries,''
\emph{Trans. AMS.},vol. 238, pp. 355-383, April 1978.




\bibitem{LC23}
S. Liu and Q. Chen,``Entropy functions on two-dimensional faces of
polymatroidal region of degree four,'' accepted by IEEE
Int. Symp. Info. Theory, Taipei, China,  June
  2023.

\bibitem{LCC24}
S.Liu, Q.Chen, and M.Cheng, “Information-theoretic constraints breed
new combinatorial structures: Entropy functions on two-dimensional
faces of polymatroidal region of degree four,” in preparation.


\bibitem{JY10}
L. Ji, J. Yin, ``Constructions of new orthogonal arrays and covering
arrays of strength three,'' \emph{J. Combin. Theory Ser. A} 117 (2010) 236-247.


\bibitem{YWJL11}
J. Yin, J. Wang, L. Ji, Y. Li, ``On the existence of orthogonal arrays
$\oa(3, 5, 4n+2)$,'' \emph{J. of Combin. Theory, Ser. A,} 118 (2011)
270–276.



\end{thebibliography}
\end{document}